\documentclass[notitlepage,onecolumn,prd,showpacs,amsmath,amssymb]{revtex4-1}
\usepackage{amssymb}
\usepackage{graphicx}
\usepackage{bm}
\usepackage{epstopdf}
\usepackage[super]{nth}
\usepackage{epsfig}
\usepackage{dsfont}
\usepackage{bigints}
\usepackage{amsmath}
\usepackage{relsize}
\usepackage{relsize}
\usepackage{amsmath}
\usepackage{accents}
\newlength{\dhatheight}

\newcommand{\unit}{1\!\!1}
\newcommand{\s}[1]{\scriptscriptstyle{#1}}

\begin{document}

\title{A quantum formalism for events and how time can emerge from its foundations}
\author{Eduardo O. Dias}
\email[]{eduardodias@df.ufpe.br}
\affiliation{Departamento de
F\'{\i}sica, Universidade Federal de Pernambuco, Recife, Pernambuco
50670-901, Brazil\\Clarendon Laboratory, Department of Physics, University of Oxford, Oxford OX1 3PU, United Kingdom}

\begin{abstract}
\emph{}
Although time is one of our most intuitive physical concepts, its understanding at the fundamental level is still an open question in physics. For instance, time in quantum mechanics and general relativity are two distinct and incompatible entities. While relativity deals with events (points in spacetime), with time being observer-dependent and dynamical, quantum mechanics describes physical systems by treating time as an independent parameter. To resolve this conflict, in this work, we extend the classical concept of an event to the quantum domain by defining an event as a transfer of information between physical systems. Then, by describing the universe from the perspective of a certain observer, we introduce quantum states of events with space-time-symmetric wave functions that predict the joint probability distribution of a measurement (observation) at $ (t, {\vec x}) $. Under these circumstances, we propose that a well-defined instant of time, like any other observable, arises from a single event, thus being an observer-dependent property. As a result, a counterfactual asymmetry along a particular sequence of events within a stationary quantum state gives rise to the flow of time as being successive ``snapshots'' from the observer's perspective. In this proposal, it is the many distinguishable states in which the observer stores information that makes the existence of time possible. 

\end{abstract}
\maketitle

\section{Introduction}
\label{intro}

In classical and quantum mechanics (QM), time is an extrinsic variable that can be chosen arbitrarily to evaluate the state of a system~\cite{Bar}. In the quantum scenario, this feature is embedded in the very definition of state, e.g., in
\begin{eqnarray}\label{stateS}
\big|\psi(t)\big\rangle&=&\sum_\alpha ~\psi(\alpha,t)~\big|\alpha\big\rangle,
\end{eqnarray}
where $|\psi(t)\rangle$ is the state of a system $\cal S$ at the instant $t$ and $\hat \alpha$ is an observable such that ${\hat \alpha}|\alpha\rangle=\alpha|\alpha\rangle$. To predict experimental outcomes, the wave function $\psi(\alpha,t) =\langle \alpha |\psi (t)\rangle $ is interpreted as the probability amplitude of measuring the system in the state $ | \alpha \rangle $, \emph{given that} the detection takes place at time $t$~\cite{Page,Vedral2,Gio,Vedral,Dias1} . To avoid any confusion with other functions defined in the present work, from now on we will highlight the time-conditioned character of the Schr\"odinger wave function with the notation $\psi(\alpha|t)$. In the orthodox interpretation of QM, because of the intrinsic uncertainty of the instant $t$ of a measurement, this time-conditioned aspect yields an indefiniteness of the moment at which the Schr\"ondiger evolution should be interrupted to calculate the probabilities of $\alpha$ via $|\psi(\alpha|t)|^2$. This dilemma is an important element of the traditional measurement problem~\cite{Zeh}. Besides that, it is broadly familiar that time being a parameter goes against the foundations of a possible quantum theory of gravity, which should treat space and time on an equal footing~\cite{Zeh2}.

There are two well-known assumptions that aim to overcome those inconsistencies: (i) in one of them, it is argued that a quantum formalism compatible with general relativity (GR) --- in which there is no preferred independent time variable --- should describe only the relative evolution between physical quantities~\cite{Is,Rov}. As we will discuss below, the Page and Wootters (PW) formalism approached in Refs.~\cite{Page,Gio,Vedral,Vedral2}, whose Hamiltonian constraint is compatible with formalisms about quantum gravity~\cite{Witt}, promptly fulfills this requirement. (ii) The second hypothesis refers to the fact that as GR deals with events (points in space-time), a suitable quantum approach should predict these events by modeling space and time symmetrically~\cite{Rov,Maccone,Rov2,Rov3,Cot}. It is worth noticing that, different from quantum states of physical systems, which extend indefinitely in time, an event has a finite temporal domain~\cite{Maccone}. For instance, a QM for events should provide, even in the non-relativistic scenario, the probability of measuring a particle in the region $ x + dx $ and the interval $ t + dt $. Note that by setting $ \alpha = x $ in the state~(\ref{stateS}), $ | \psi (t) \rangle $ does not carry that information. Nevertheless, despite the coherence of the conditions (i) and (ii), the attempt to reconcile at a fundamental level these two criteria [more specifically, the PW picture and the case (ii)] requires special attention if one intends to extend the classical notion of events to the quantum domain. Let us address this standpoint below.

In the PW picture, it is wisely assumed that the universe is stationary, and time arises from the correlation between a system $\cal C$  that plays the role of a clock and the rest of the universe $\cal R$~\cite{Vedral2}. The clock is conveniently chosen not to interact with the rest so that $\cal R$ evolves according to the Schr\"odinger equation with respect to a variable $\beta$ that labels the eigenvalues of a proper clock's observable. For instance, in a universe with only two degrees of freedom, e.g., $\alpha$ e $\beta$, the rest $\cal R$ has a $\beta$-conditioned wave function $ \psi (\alpha | \beta)$ given by Eq.~(\ref{stateS}), with $ t=\beta $. Under these circumstances, $\beta$ plays the same role as $t$ in the Schrodinger prescription~(\ref{stateS}), and thus time is no longer merely a parameter, but rather an observable that tracks the dynamics of $\cal R$. It is worth mentioning that in this manuscript, we will propose a different way of looking at the emergence of time. In a classical version of the PW picture, Newton's laws also provide a ``timeless'' relationship between $\alpha$ and $\beta$, such as $F(\alpha,\beta)=0$. Therefore, by conveniently isolating $\alpha$ in $F(\alpha,\beta)=0$, we have either the classical and quantum versions of the relative evolution between observables required in the criterion (i) above,
\begin{eqnarray}\label{corre}
\alpha=\alpha(\beta)~~\big({\rm classical~state}\big)~~~~~~{\rm and}~~~~~~\psi=\psi(\alpha|\beta)~~\big({\rm quantum~state}\big).
\end{eqnarray}
For a given $\beta$, classically, we have a well-defined value of $ \alpha$, whereas, in the quantum scenario, we only have a probability amplitude for $\alpha$.

We now turn our attention to the description of events in the PW approach. An event is a happening, for example, a classical particle $ \cal S $ (belonging to $\cal R$) with position $\alpha$ reaches a mark on the surface on which it is moving. In the classical regime, as we know with arbitrary precision the reading of the clock at the arrival moment ($\beta=t$) of the particle, by applying the first expression of Eq.~(\ref{corre}), the event is fully described by the pair $ (t, \alpha (t))$. In contrast, in the quantum domain, as both $\alpha$ and the arrival time $t$ are probabilistic variables~\cite{Muga,All,Kij,Hal2,Bau}, an event should be depicted by a joint probability amplitude of $t$ and $\alpha(t)$. Thus, the classical and quantum descriptions of events, differently from Eq.~(\ref{corre}), should be given by
\begin{eqnarray}\label{correq}
(t, \alpha (t))~~\big({\rm classical~event}\big)~~~~~~{\rm and}~~~~~~\Psi(t,\alpha)=\psi(\alpha|t)~\chi(t)~~\big({\rm quantum~event}\big),
\end{eqnarray}
with $|\Psi(t,\alpha)|^2$ being the joint probability of the event $(t,\alpha(t))$.  By applying Bayes' rule, $ | \Psi (t, \alpha) |^ 2 $ in Eq.~(\ref{correq}) is expressed as the probability of measuring $ \alpha $ given that the clock reads $ \beta = t $, $|\psi(\alpha|t)|^2$, multiplied by the probability for the clock to read $t$ at the moment of the event, $|\chi(t)|^2$. Here, notice that we need a detector $\cal D$ (also belonging to $\cal R$) that, by measuring the particle $\cal S$, defines its arrival. It is worth mentioning that because of the measurement interaction, $\psi(\alpha|t)$ in Eq.~(\ref{correq}) is, in general, different from that of the isolated situation of Eq.~(\ref{stateS}).

In this description of events, although the time probability amplitude $\chi(t)$ concerns the reading of the clock $\cal C$ at the moment of the arrival, as $\cal C$ is a non-interacting system in the PW picture~\cite{Gio,Vedral,Vedral2}, $\chi(t)$ cannot be a wave function of this clock. For such a link to be possible, for instance, $\cal C$ should be coupled to the detector $\cal D$, which defines the arrival of the particle. Then, another clock, non-interacting with $\cal R$, should be taken into account to maintain the PW approach valid. Hence, $\cal R$ would still follow the Schr\"odinger equation, but now with respect to an appropriate observable of this new clock. As a consequence, a quantum description of events [criterion (ii)] as proposed in Eq.~(\ref{correq}) cannot be derived in the PW formalism simply through the correlation~(\ref{corre}) between observables [criterion (i)] with $t$ in $\Psi(t,\alpha)$ being a variable of the isolated clock.

In the present work, we intend to solve these issues by keeping $\cal C$ isolated in the PW approach and assuming that an event is a register of information about $\cal S$ performed within the rest $\cal R$ by a detector $\cal D$ and a timer $\cal T$. First, the timer will be modelled by a Salecker-Wigner-Peres clock~\cite{SaWi,Pe}, but, instead of being coupled with the system under consideration~\cite{Pe}, $\cal T$ will interact with the detector. This coupling will be essential in the derivation of our formalism, especially in the case of multiple events. We will verify that quantum states predicting events are not represented at a specific instant of ``time'' (a variable that tracks the dynamics) as in Eq.~(\ref{stateS}), but rather by a timeless state like
\begin{eqnarray}\label{state1}
\big|\big|\Psi\big\rangle&=&\sum_{t,\alpha}~\Psi(t,\alpha)~\big|\big|t,\alpha\big\rangle,
\end{eqnarray}
with $t$ being a variable of the timer $\cal T$, $\alpha$ the eigenvalue of a certain observable of $\cal S$, and $\Psi(t,\alpha)$ the wave function of a single event, as depicted in Eq.~(\ref{correq}). In Eq.~(\ref{state1}), $||\Psi\rangle$ will be obtained via a Wheeler-DeWitt equation~\cite{Page,Gio,Vedral,Vedral2,Witt} and, as will be carefully defined, the two bars of $||t,\alpha\rangle$ will represent the quantum state of an event, which describes $\cal CS$ from the perspective of $\cal TD$. In particular aspects, the standpoint adopted here is in line with the ``relative state'' formulation of QM~\cite{Eveint}. Nevertheless, it is worth mentioning that our approach does not require the many-worlds (or minds) view. We will also generalize $||\Psi\rangle$ for an indefinite number of causally connected events without assuming any collapse. Situations involving causal and non-causal events will be briefly discussed.

Relevant works that describe events commonly consider several ``instanteanous'' measurements~\cite{Rov3,Cot,Vedral3}. As a result, these formalisms do not provide a state for a \emph{single event} that treats time and any other observable $ \alpha $ on an equal footing as in Eq.~(\ref{state1}). On the other hand, works such as Refs~\cite{Rov2,Hal2}, which do consider measurements with a finite duration, do not take into account the recording of the measurement time, and hence Eq.~(\ref{state1}) is not applicable either. It is also important to stress that regarding the example of the particle reaching a point in space,  there is a huge body of interesting literature about arrival times that aim to obtain time probability distribution such as $\chi(t)$~\cite{Muga,All,Kij,Hal2,Bau}. Nevertheless, unlike the physical scheme of Eq.~(\ref{state1}), these works consist mostly of calculating the time of arrival either by using the wave function of a particle in the absence of measurement (i.e., the arrival time as a property of the particle), or introducing only a detector that by interacting with the particle defines its arrival. In addition, it should be remarked that Refs.~\cite{Dias1,Xim} propose the state of Eq.~(\ref{state1}) for the particular case $ \alpha = x$. However, neither the concept of events nor how to obtain $||\Psi\rangle$ from traditional tools of QM was explored in this reference. Besides, it is worth mentioning that even for $\alpha=x$, Eq.~(\ref{state1}) is a simplified version of the states we will obtain in the next sections. Nevertheless, we believe that the physical connections between the results of the present manuscript and those of Ref.~\cite{Dias1}  deserve to be investigated in future works.

The second focus of our approach is to analyse the possibility of obtaining from $||\Psi\rangle$ our familiar notion of time with a preferred flow direction. In the PW formalism, $\beta$ is called time mainly because $\psi(\alpha|\beta)$ follows the Schr\"odinger equation. Nevertheless, note that this treatment gives rise to the concept of the Newtonian time, in the sense that time can be arbitrarily chosen to assess the state of the rest $\cal R$. Therefore, it is expected that this notion of time leads to inconsistencies with our daily experience. Two emblematic examples are the incompatibility between the second law of thermodynamics and the $\beta$-reversal evolution of $\psi(\alpha|\beta)$, and the unnatural interpretation in which the reality we understand is fixed to a single $\beta$ (an ``instant of time''). From this latter viewpoint, the passage of time is a mere illusion that can arise from our perception of memories. In contrast to this approach, here we intend to address a concept of time more congruent with our common sense. 

From another perspective of the problem of time, several works such as Refs.~\cite{Linden,Short,Reimann} found that as time progresses, the increasing entanglement of an object with their surroundings makes this object reaches equilibrium. From this result, the arrow of time is commonly associated with an increase of correlations. However, it should be pointed out that this approach does not explain the nature of time itself, but rather the reason why physical systems reach equilibrium over time. The question of why we perceive time flowing is an open issue that we intend to address in this work. To this end, first, we will verify that a more fundamental description of events should take into account that an observer with many degrees of freedom can play the same role as a set of timers $\cal T$. Then, by proposing that time emerges from events, we will observe that the flow of time arises from the perspective of the observer as a consequence of a causal-like asymmetric sequence of events within $||\Psi\rangle$. Here, causal-like means a counterfactual conditional~\cite{Good} such that without the need for any relationship with time, in a ``selection'' of an indefinite number of variables from a set $\{X_{  (k)}\}_{k=1,2,3,\hdots}$, if $X_{ (k)}$ is not selected, $X_{(k+1)}$ cannot be selected.

A latter investigation concerns the usual assumption in which, during a measurement process, particles do not interact except with the measuring device~\cite{Biru}. This simplification only works well when the time of observation is much shorter compared to the interaction time. In this picture, we know the detection time $t$, and thus we are able to properly choose $\beta=t$ so that $|\psi(t)\rangle$ in Eq.~(\ref{stateS}) can predict experimental results accurately. However, when the time scale of the measurement is not small enough and the detection time is somehow registered, we will verify that experimental outcomes should follow predictions from wave functions like $\Psi(\alpha,t)$ in Eq.~(\ref{correq}), as expected.

This manuscript is organized as follows. In Sec.~\ref{form}, we will describe events by conditioning $ || \Psi \rangle $ on a specific $ \beta $ of the clock $\cal C$. In Sec.~\ref{form1e}, the case of a single event will be discussed, and then in Sec.~\ref{form2e} we will extend this analysis to two causally connected events. In Sec.~\ref{env}, we will verify that a system with many degrees of freedom can play the role of the timer $\cal T$. In Sec.~\ref{qme}, we will formulate the quantum description of events via a steady-state $ || \Psi \rangle $. In Secs.~\ref{qme1} and~\ref{timelike}, the case of a single event and an arbitrary number of causally connected events will be approached respectively. In Sec.~\ref {time}, we propose how time can emerge from $ || \Psi \rangle $. Lastly, the generalization for causally and non-causally connected events will be briefly addressed in Sec.~\ref{spacelike}. In the conclusion of this manuscript, Sec.~\ref{conclusion}, we will summarize the results.

\section{Events from the perspective of conditioned quantum states}
\label{form}

\subsection{A single event}
\label{form1e}

Let us begin this section by briefly reviewing the PW formalism~\cite{Page,Gio,Vedral2}, with the clock $\cal C$ and the rest of the universe $ \cal R $ having Hilbert spaces $ {\cal H}_{\cal C} $ and ${\cal H}_{\cal R}$ respectively. It should be clear that, as we will only discuss the emergence of time in Sec.~\ref{time}, for the sake of understanding, we will frequently refer to the clock reading $\beta$ as a measure of time. Thus, for convenience, let us define the possible values of $\beta = t_0, t_{\s (1)}, t_{\s (2)}, \hdots, t_{\s (N)}, \hdots$ . The condition of an isolated universe implies that $||\Psi\rangle$ belonging to ${\cal H}_{\cal C} \otimes {\cal H}_{\cal R}$ must satisfy the Wheeler-DeWitt constraint~\cite{Page,Vedral2,Gio,Witt}
\begin{eqnarray}\label{PW}
{\hat H}\big|\big|\Psi \big\rangle=0,
\end{eqnarray}
where
\begin{eqnarray}\label{PWH}
{\hat H}={\hat H}_{\cal C}\otimes \unit_{\cal R} + \unit_{\cal C} \otimes {\hat H}_{\cal R}.
\end{eqnarray}
By considering $\hat \beta$, with ${\hat \beta}|\beta\rangle_{\cal C}=\beta|\beta\rangle_{\cal C}$, conjugated to the clock’s Hamiltonian ${\hat H}_{\cal C}$, $[{\hat \beta},{\hat H}_{\cal C}]=i\hbar$, the static solution of Eq.~(\ref{PW}) can be written as
\begin{eqnarray}\label{PWsol}
\big|\big|\Psi\big\rangle=\sum_{\beta=t_0}~ \big|\beta\big\rangle_{\cal C}\otimes\big|\psi(\beta)\big\rangle_{\cal R},
\end{eqnarray}
with $|\beta\rangle_{\cal C}=\exp\big\{-i{\hat H}_{\cal C}(\beta-t_0)/\hbar\big\}|t_0\rangle_{\cal C}$ and
\begin{eqnarray}\label{Schr}
\big|\psi(\beta)\big\rangle_{\cal R}={\rm e}^{-i{\hat H}_{\cal R}(\beta-t_0)/\hbar}~\big|\psi(t_0)\big\rangle_{\cal R}.
\end{eqnarray}
We choose the state $|t_0\rangle_{\cal C}$ of the clock to indicate the beginning of the evolution. Notice that the Schr\"odinger state of $\cal R$ is obtained by conditioning $||\Psi\rangle$ on $\beta$, i.e. $|\psi(\beta)\rangle_{\cal R}={_{\cal C}\langle} \beta||\Psi\rangle \in {\cal H}_{\cal R}$. We will first describe events via the conditioned state $|\psi(\beta)\rangle_{\cal R}$, but it is worth mentioning that the formalism will only be fully elaborated in the next sections when we will deal with $||\Psi\rangle$.

An event will be treated as an effectively irreversible measurement carried out within $\cal R$, in which an observer records information about a system $\cal S$ with Hamiltonian ${\hat H}_{\cal S}$. Consider the unitary evolution~(\ref{Schr}) during the measurement of an observable ${\hat \alpha}=\sum_\alpha\alpha|\alpha\rangle_{\cal S}{_{\cal S} \langle \alpha|}$ of $\cal S$ that is performed by a detector $\cal D$ (the observer). For instance, we can assume that either $\cal D$ is coupled to a large environment $\cal E$, which makes the detection effectively
irreversible~\cite{Hal2}, or $\cal D$ is a macroscopic system since, in realistic measurements, detectors have a huge number of degrees of freedom. For now, we disregard the recording of the instant of detection (there is no timers), and consider that the interaction time $\Delta t_{\cal SD}$ between the detector and the system is short enough to neglect the evolution driven by ${\hat H}_{\cal S}$.

Under these circumstances, with $\cal R=SD$, Eq.~(\ref{PWH}) becomes ${\hat H}\approx{\hat V}_{\cal SD}\otimes \unit_{\cal C} + \unit_{\cal SD} \otimes {\hat H}_{\cal C}$, where ${\hat V}_{\cal SD}$ is the interaction potential between $\cal S$ and $\cal D$. The solution~(\ref{Schr}) with initial condition $|\psi(t_0)\rangle_{\cal R}=|0\rangle_{\cal S}\otimes|0\rangle_{\cal D}$ is
\begin{eqnarray}\label{VN}
&~&\big|0\big\rangle_{\cal S}\otimes\big|0\big\rangle_{\cal D}=\bigg( \sum_\alpha~ \psi_{\cal S}(\alpha|t_0)~\big|\alpha\big\rangle_{\cal S} \bigg)\otimes \big|0\big\rangle_{\cal D}~~\underrightarrow{~~\Delta t_{\cal SD}~~}~~
\sum_\alpha~ \psi_{\cal S}(\alpha|t_0)~\big|\alpha\big\rangle_{\cal S}\otimes \big|\alpha\big\rangle_{\cal D}=\sum_\alpha ~{\hat M}_\alpha~ \big|0\big\rangle_{\cal S}\otimes \big|\alpha \big\rangle_{\cal D},
\end{eqnarray}
where $|0\rangle_{\cal D}$ is the ready state of the detector, $|0\rangle_{\cal S}=|\psi(t_0)\rangle_{\cal S}$, $\psi_{\cal S}(\alpha|t_0)=\langle \alpha|0\rangle_{\cal S}$, and  ${\hat  M}_\alpha=|\alpha\rangle_{\cal S}{_{\cal S} \langle \alpha|}$. As mentioned earlier, we use the notation $\psi_{\cal S}(\alpha|t_0)$ instead of the usual nomenclature, such as $c_\alpha(t_0)$, to emphasize the time-conditioned character of the Schr\"odinger amplitudes. In Eq.~(\ref{VN}), although we neglected the evolution of $\cal S$ by ${\hat H}_{\cal S}$, it is important to keep in mind that $\Delta t_{\cal SD}$ is actually finite. Thus, notice that Eq.~(\ref{VN}) guarantees that by observing the detector at some moment $\beta \gtrsim t_0 + \Delta t_{ \cal SD}$, we unequivocally find out the state $|\alpha\rangle_{\cal S}$. Nevertheless, as will be more evident in Sec.~\ref{env}, if $\cal D$ had only one degree of freedom to store information about $\cal S$, one could not claim that the detector measured $\cal S$ at some instant $t$ before the observation of $\cal D$, i.e., in the interval $t_0<t< \beta$. The reason for this is that before $ \beta \approx t_0 + \Delta t_{\cal SD} $, the pair ${\cal SD}$ is in a superposition that only describes the unmeasured (detector in $ | 0 \rangle_{\cal D }$) and measured (detector in $|\alpha \rangle_{\cal D} $) situations. As a consequence, unlike the classical picture, the time of a quantum event (or detection) only exists if a physical system somehow registers it.

With these facts in mind, we begin our approach describing the classical event $(t,\alpha(t))$ in the quantum domain by keeping the clock $\cal C$ isolated and, first, using a SWP's-like timer $\cal T$~\cite{SaWi,Pe} synchronized with $\cal C$. We will not yet analyse the possibility of $\cal D$ (a macroscopic system) recording the instant of detection. This more general description, in which $\cal T$ can be disregarded, will be presented in Sec.~\ref{env}. From now on, $\Delta t_{\cal SD}$ will not be short enough to neglect the evolution driven by ${\hat H}_{\cal S}$. Under these circumstances, remember that $\chi(t)$ in Eq.~(\ref{correq}) is the probability amplitude for the clock $\cal C$ to read $t$ at the moment of the event. To obtain such an amplitude, let us couple the timer with the detector in such a way that $\cal T$ stops its counting when $\cal D$ measures $ \cal S$. Thus, $ \cal T $ evolves synchronized to $ \cal C $ while $\cal D$ ``is'' in the state $ |0 \rangle_{\cal D}$. For simplicity, consider that $\cal T$ ``instantly'' recognizes the detection of $ \cal S $, which means that the time scale of the interaction between the timer and the detector is significantly smaller than $\Delta t_{\cal SD}$. In this approximate picture, the probability amplitude of $\cal T$, $\chi(t)$ [see Eq.~(\ref{correq})], predicts the reading of $\cal C$ at the moment of the event. Now, the observer will be seen as $\cal TD$. It is worth mentioning that despite the continuous monitoring of the timer, the detector is not affected by the Zeno effect as long as it is macroscopic~\cite{Zeh}.

An interaction between $\cal T$ and $\cal D$ that models the evolution defined above is ${\hat V}_{\cal TD}=\unit_{\cal S} \otimes {\hat H}_{\cal T}\otimes |0\rangle_{\s \cal D}{_{\s \cal D}\langle} 0|$~\cite{Pe}. As every measurement is carried out internally to $\cal R$, we have
\begin{eqnarray}\label{hamilt}
{\hat H}={\hat H}_{\cal C}\otimes \unit_{\cal STD} + \unit_{\cal C}\otimes {\hat H}_{\cal STD},
\end{eqnarray}
where ${\hat H}_{\cal STD}={\hat H}_{\cal S}\otimes\unit_{\cal TD}+{\hat V}_{\cal SD}+  {\hat V}_{\cal TD}$. Since Eq.~(\ref {hamilt}) has the same form as the PW's Hamiltonian~(\ref{PWH}), the solutions~(\ref{PWsol}) and~(\ref{Schr}) still hold. Let us consider the timer's observable $\hat T$ such that ${\hat T}|t\rangle_{\cal T}=t|t\rangle_{\cal T}$ and $[{\hat T},{\hat H}_{\cal T}]=i\hbar$, and the initial condition of $\cal R=\cal STD$ being
\begin{eqnarray}\label{initial}
\big|\psi(t_0)\big\rangle_{\cal R}=\big|0\big\rangle_{\cal S}\otimes\big|t_0\big\rangle_{\cal T}\otimes\big|0\big\rangle_{\cal D}.
\end{eqnarray}
To prevent a dense notation, from now on, we will avoid using the symbol $\otimes$.

For the sake of clarity, let us consider $\cal S$ being composed of a single system. To calculate $|\psi(\beta)\rangle_{\cal R}$, let us break up the evolution~(\ref{Schr}) into infinitesimal steps $\delta t=t_{\s (j+1)}-t_{\s (j)} \ll \Delta t_{ \cal SD}$, so that the possible values of $\beta$ and $t$ in this discrete evolution are $t_0, t_{\s (1)}, t_{\s (2)}, \hdots, t_{\s (N)}, \hdots$. At the first step $t_{\s (1)}$, the state of $\cal R=STD$ splits into two branches (system $\cal S$ measured and not measured) given by
\begin{eqnarray}
\label{inf1}
\big|\psi(t_{\s (1)})\big\rangle_{\cal R}=\sqrt{1-\delta p_{\s (1)}}~{\rm e}^{i\varphi_{\s (1)}}~\bigg[{\hat U}_{{\cal S}}^{  \s 0(1)}(t_{\s (1)},t_0)\big|0\big\rangle_{\cal S}\bigg]~\big|t_{\s (1)}\big\rangle_{\cal T}~\big|0\big\rangle_{\cal D}~+~\sqrt{\delta p_{\s (1)}}~\bigg[\sum_\alpha{\hat M}_\alpha~{\hat U}_{\cal S}^{  \s m(1)}(t_{\s (1)},t_0)\big|0\big\rangle_{\cal S}\bigg]~\big|t_{\s (1)}\big\rangle_{\cal T}~\big|\alpha\big\rangle_{\cal D},\nonumber\\
\end{eqnarray}
where $\varphi_{\s(1)}$ is the phase of the first step, and $\delta p_{\s (1)}\ll 1$ is the probability of the measurement taking place in the interval $[t_0,t_{\s (1)}]$, regardless of the outcome $|\alpha\rangle_{\cal S}$. Here, the operations ${\hat U}_{\cal S}^{\s 0(1)}(t_{\s (1)},t_0)$ and ${\hat U}^{\s m(1)}_{\cal S}(t_{\s (1)},t_0)$ act on ${\cal H}_{\cal S}$ and refer to the situations where $\cal S$ is not measured and measured in the first step, respectively. The interaction between $\cal S$ and $\cal D$ makes these operations, in general, different from ${\hat U}_{\cal S}(t_{\s (1)},t_0)=\exp\{-i{\hat H}_{\cal S}(t_{\s (1)}-t_0)/\hbar\}$. For instance, if for each $|\alpha\rangle_{\cal S}$, the detector interacts with a different intensity, ${\hat U}_{\cal S}^{\s 0(1)}(t_{\s (1)},t_0)$ and ${\hat U}^{\s m(1)}_{\cal S}(t_{\s (1)},t_0)$ are no longer unitary [see the appendix for a brief discussion of these operations and Eq.~(\ref{inf1})]. Nevertheless, the total evolution obviously is always unitary, so that $_{\cal R}\langle\psi(t_{\s(1)})|\psi(t_{\s(1)})\rangle_{\cal R}=1$. All the quantities of Eq.~(\ref{inf1}) can be calculated via Eq.~(\ref{Schr}) by appropriately defining the potentials of ${\hat H}_{\cal STD}$. By inspecting Eq.~(\ref{inf1}), first we verify that with a high probability $1-\delta p_{\s (1)}$, the detector does not record any information about the system. In this case, the timer continues to evolve as an ideal quantum clock. On the other hand, with probability $\delta p_{\s (1)}$, the detector measures some state $|\alpha\rangle_{\cal S}$, and hence the timer stops its evolution recording $t_{\s (1)}$, which is the reading of $ \cal C $ at this moment.

In the next step of the evolution, the unmeasured contribution of $|\psi(t_{\s (1)})\rangle_{\cal R}$ splits into two branches, so that
\begin{eqnarray}\label{psi2}
\big|\psi(t_{\s (2)})\big\rangle_{\cal R}&=&\sqrt{1-\delta p_{\s (1)}}~{\rm e}^{i\varphi_{\s (1)}}~\sqrt{1-\delta p_{\s (2)}}~{\rm e}^{i\varphi_{\s (2)}}~\bigg[{\hat U}_{{\cal S} }^{ \s 0(2)}(t_{\s (2)},t_{(1)})~{\hat U}_{{\cal S} }^{ \s 0(1)}(t_{\s (1)},t_0)\big|0\big\rangle_{\cal S}\bigg]~\big|t_{\s (2)}\big\rangle_{\cal T}~\big|0\big\rangle_{\cal D}\nonumber\\
~\nonumber\\
&+&\sqrt{1-\delta p_{\s (1)}}~{\rm e}^{i\varphi_{\s (1)}}~\sqrt{\delta p_{\s (2)}}~\bigg[\sum_\alpha{\hat M}_\alpha~{\hat U}_{\cal S}^{\s m(2)}(t_{\s (2)},t_{\s(1)})~{\hat U}_{\cal S}^{\s 0(1)}(t_{\s (1)},t_0)~\big|0\big\rangle_{\cal S}\bigg]~\big|t_{\s (2)}\big\rangle_{\cal T}~\big|\alpha\big\rangle_{\cal D}\nonumber\\
~\nonumber\\
&+&\sqrt{\delta p_{\s (1)}}~\bigg[{\hat U}_{\cal S}(t_{\s (2)},t_{\s (1)})~\sum_\alpha {\hat M}_\alpha~{\hat U}_{\cal S}^{\s m(1)}(t_{\s (1)},t_0)\big|0\big\rangle_{\cal S}\bigg]~\big|t_{\s (1)}\big\rangle_{\cal T}~\big|\alpha\big\rangle_{\cal D}.
\end{eqnarray}
By inspecting the first term of Eq.~(\ref{psi2}), we verify that the probability for the system to remain unmeasured until $t_{\s (2)}$ is $ (1-\delta p_{\s (1)}) (1- \delta p_{\s (2)})$, and from the the second term, we observe that $ \cal S $ is measured in the second step with probability $ (1- \delta p_{\s (1 )}) \delta p_ {\s (2)} $. Finally, in the last branch of $ | \psi (t_{\s (2)}) \rangle_{\cal R} $, as the measurement happened at the previous instant $ t_{\s (1)}$ and $\cal S$ is no longer under measurement, the information remains recorded in $\cal TD$, and $\cal S$ evolves according to ${\hat U}_{\cal S}(t_{\s (2)},t_{\s (1)})$.

Keeping this process up to the $N$-th step, we have
\begin{eqnarray}\label{infN}
\big|\psi(t_{\s (N)})\big\rangle_{\cal R}&=&\Gamma(t_{\s (N)},t_0)~\bigg[{\hat U}^{\s 0}_{\cal S}(t_{\s (N)},t_0)\big|0\big\rangle_{\cal S}\bigg]~\big|t_{\s (N)}\big\rangle_{\cal T}~\big|0\big\rangle_{\cal D}\nonumber\\
~\nonumber\\
&+&~\sum_{k=1}^{N} ~\chi(t_{\s (k)})~\bigg[{\hat U}_{\cal S}(t_{\s (N)},t_{\s (k)})~\sum_\alpha{\hat M}_\alpha~{\hat U}^{\s (k)}_{\cal S}(t_{\s (k)},t_0)\big|0\big\rangle_{\cal S}~\bigg]~\big|t_{\s (k)}\big\rangle_{\cal T}~\big|\alpha\big\rangle_{\cal D},
\end{eqnarray}
where

\begin{eqnarray}\label{U0}
{\hat U}^{\s 0}_{\cal S}(t_{\s (k)},t_0)={\hat U}^{\s 0(k)}_{\cal S}(t_{\s (k)},t_{\s (k-1)})~{\hat U}^{\s 0(k-1)}_{\cal S}(t_{\s (k-1)},t_{\s(k-2)})\hdots{\hat U}^{\s 0(1)}_{\cal S}(t_{\s (1)},t_0),
\end{eqnarray}
\begin{eqnarray}\label{Um}
{\hat U}^{\s (k)}_{\cal S}(t_{\s (k)},t_0)={\hat U}^{\s m(k)}_{\cal S}(t_{\s (k)},t_{\s (k-1)})~{\hat U}^{\s 0(k-1)}_{\cal S}(t_{\s (k-1)},t_{\s(k-2)})\hdots{\hat U}^{\s 0(1)}_{\cal S}(t_{\s (1)},t_0),
\end{eqnarray}
\begin{eqnarray}\label{Gamma}
\Gamma(t_{\s N},t_0)=\prod_{k=1}^{{N}}\sqrt{1-\delta p_{\s (k)}}~{\rm e}^{i\varphi_{\s (k)}}~~~{\rm and}~~~\chi(t_{\s (k)})= \sqrt{\delta p_{\s (k)}}~\prod_{\ell=1}^{k-1}\sqrt{1-\delta p_{\s (\ell)}}~{\rm e}^{i\varphi_{\s (\ell)}}.
\end{eqnarray}
Here, $|\Gamma(t_{\s N},t_0)|^2$ is the probability of the system not being measured from the beginning of the process until $t_{\s (N)}$, and accordingly, $\cal D$ has no information about $\cal S$ in this branch of Eq.~(\ref{infN}). On the other hand, in the second expression of $|\psi(t_{\s (N)})\rangle_{\cal R}$, for a given value of $k$, the measurement took place in the interval $[t_{\s (k-1)},t_{\s (k)}]$ with probability $|\chi(t_{\s (k)})|^2$, regardless of the outcome $|\alpha\rangle_{\cal S}$. Let us call $\chi(t_{\s (k)})$ the probability amplitude of the clock reading $t_{\s (k)}$ at the moment of the event. It should be noted that the width of $\chi(t_{\s (k)})$ is of the order of $\Delta t_{\cal SD}$. Thus, as all states $\{|\alpha\rangle_{\cal S}\}$ are under measurement, the probability of the system not being measured is negligible when $\beta=t_{\s (N)} \gtrsim t_0+\Delta t_{\cal SD}$, and hence $\Gamma(t_{\s (N)})$  and $\chi(t_{\s (N)})$ are $\approx 0$ in Eq.~(\ref{infN}). Unlike the ideal measurement~(\ref{VN}), the recording of the detection time by $\cal T$ allows us to claim that the event in fact occurred at some instant $t_{\s (k)}$ in the interval $ [t_0,t_0+\Delta t_{ \cal SD}] $. It is worth pointing out that if $\cal S$ is composed of many indepedent subsystems, we can obtain the same expression as Eq.~(\ref{infN}) as long as we do not specify the subsystem of $\cal S$ that is measured by $ {\cal D} $. For instance, if $\cal S$ is composed of non-interacting and distinguishable particles ${\cal S}_i$ (with $i=1,2,\hdots$), and $\cal D$ measures one subsystem at a time, the same calculation above also results in Eq.~(\ref{infN}), but with ${\hat M}_\alpha=\sum_i{\hat M}_{S_i,\alpha}$, where ${\hat M}_{S_i,\alpha}=\unit_{S_1}\otimes\unit_{S_2}\otimes\hdots\otimes{\hat M}_{\alpha_i}\otimes\unit_{S_{i+1}} \hdots$. Situations involving entangled systems is the subject of a work in progress.

Let us verify that $|\psi(t_{\s (N)})\rangle_{\cal R}$ provides the same information as $||\Psi\rangle$ presented in Eq.~(\ref{state1}) of the introduction. By calling $t=t_{\s (k)}$ and rewriting the sum  over $k$ as a sum over $t$, Eq.~(\ref{infN}) evaluated after one guarantees that the measurement takes place [$\beta=t_{\s (N)} \gtrsim t_0+\Delta t_{\cal SD}$] becomes
\begin{eqnarray}\label{infN2}
\big|\psi(t_{\s (N)})\big\rangle_{\cal R}=\sum_{\alpha, t> t_0}^{t_{\s (N)}}~\psi_{\cal S}(\alpha|t)~\chi(t)~\bigg[{\hat U}_{\cal S}(t_{\s (N)},t)~\big|\alpha\big\rangle_{\cal S}~\big|t,\alpha \big\rangle_{\cal TD}\bigg],
\end{eqnarray}
where $|t,\alpha\rangle_{\cal TD}=|t\rangle_{\cal T}|\alpha\rangle_{\cal D}$, and the normalization conditions are
\begin{eqnarray}\label{norm}
\sum_{\alpha, t> t_0}^{t_{\s (N)}}~ |\psi_{\cal S}(\alpha|t)|^2~|\chi(t)|^2=1 ~~~~~{\rm and}~~~~~ \sum_{t> t_0}^{t_{\s (N)}} ~|\chi(t)|^2=1.
\end{eqnarray}
In Eq.~(\ref{infN2}), the expression within the brackets was obtained by rewriting the state of $\cal S$ in the second branch of Eq.~(\ref{infN}) as
\begin{eqnarray}\label{coef1}
{\hat M}_\alpha~{\hat U}_{\cal S}^{\s (k)}(t_{\s (k)},t_0)~\big|0\big\rangle_{\cal S}=\psi_{\cal S}(\alpha|t_{\s (k)})~\big|\alpha\big\rangle_{\cal S},~~~{\rm with}~~~\psi_{\cal S}(\alpha|t_{\s (k)})={_{\cal S}\big\langle}\alpha\big|{\hat U}^{\s (k)}_{\cal S}(t_{\s (k)},t_0)\big|0\big\rangle_{\cal S}
\end{eqnarray}
being the Schr\"odinger ``wave function'' (amplitude) of the observable $\hat \alpha$ of $\cal S$. Notice that Eq.~(\ref{infN2}) is a superposition of the ideal measurements~(\ref{VN}), with each branch describing a different instant of detection. In practice, Eq.~(\ref{infN2}) can be applied to predict outcomes obtained by an experimentalist that verifies the recordings of $ \cal TD $ when the clock $ \cal C $ reads $ \beta= t_{\s (N)} \gtrsim t_0 + \Delta t_{\cal SD} $. It is noteworthy that a more rigorous analysis of this problem would require the addition, for instance, of the experimentalist's brain developing the role of a detector to the theoretical calculations. Nevertheless, in this more elaborated approach, the same predictions as Eq.~(\ref{infN2}) should be obtained by conditioning  $ || \Psi \rangle $ on her brain. The reason for this agreement is that, as the information about $\cal CS$ is already recorded in $ \cal TD $ before the interaction with the experimentalist, the probabilities of the event $(t,\alpha)$ would not change if we added her to the evolution~(\ref{Schr}).

Under such circumstances, given that the experimentalist observes $\cal TD$ when $\beta=t_{\s (N)} \gtrsim t_0+\Delta t_{\cal SD}$, the probability for her to verify that the detector measured $\cal S$ in $|\alpha\rangle_{\cal S}$ with the clock $\cal C$ reading $t$ is
\begin{eqnarray}\label{prob}
P(t,\alpha)=\Big|{_{\cal TD}\big\langle} t,\alpha\big|\psi(t_{\s (N)})\big\rangle_{\cal R}\Big|^2=\big|\psi_{\cal S}(\alpha|t)\big|^2~\big|\chi(t)\big|^2.
\end{eqnarray}
This probability is the modulus squared of the amplitude (\ref{correq}) addressed in the introduction: the probability of measuring $ \alpha $ given that the clock reads $ \beta = t $, $|\psi(\alpha|t)|^2$, multiplied by the probability for the clock to read $t$ at the moment of the event, $|\chi(t)|^2$. In Eq.~(\ref{infN2}), as ${\hat U}_{\cal S}(t_{\s (N)},t)$ only acts on $\cal S$, it does not influence the probability distribution~(\ref{prob}). Also, the probability of measuring $ | \alpha \rangle_{\cal S} $ regardless of the instant of detection can be calculated as
\begin{eqnarray}\label{prob2}
P(\alpha)={\rm Tr}\Big\{\big|\psi(t_{\s (N)}) {\big\rangle_{\cal R}}_{\cal R}{\big\langle} \psi(t_{\s (N)})\big|\otimes\big|\alpha{\big \rangle_{\cal D}} _{\cal D}{ \big \langle} \alpha\big|\Big\}=\sum_{t>t_0}~P(t,\alpha)=\sum_{t>t_0}~\big|\psi_{\cal S}(\alpha|t)\big|^2~\big|\chi(t)\big|^2.
\end{eqnarray}
Because $\chi(t)\approx 0$ when $t\approx t_0+\Delta t_{\cal SD} < t_{\s (N)}$, $P(\alpha)$ does not depend on $t_{\s (N)}$, and thus the sum over $t$ can be extended to infinity. The dependence of $P(\alpha)$ on the measurement interaction is verified in both the time probability amplitude $\chi(t)$ and $\psi_{\cal S}(\alpha|t)$. Nevertheless, if the measurement is ``instantaneous'' as in the ideal case~(\ref{VN}) [$\chi(t)$ with negligible width], we can take $\psi_{\cal S}(\alpha|t)\approx \psi_{\cal S}(\alpha|t_0)$, and hence the probability of observing $\alpha$ in Eq.~(\ref{prob2}) becomes $P(\alpha)\approx|\psi_{\cal S}(\alpha|t_0)|^2$, as expected. Here, it should be used that $\sum_{t>t_0} |\chi(t)|^2=1$.

Lastly, let us turn our attention to the state of $ \cal R $ when the clock reads an arbitrary value of $ \beta $, as illustrated in Eq.~(\ref{infN}). By using the notation of Eq.~(\ref{infN2}), $|\psi(\beta)\rangle_{\cal R}={_{\cal C}\langle} \beta||\Psi\rangle$ becomes
\begin{eqnarray}\label{Psi}
\big|\psi(\beta)\big\rangle_{\cal R}=\Gamma(\beta,t_0)~\bigg[{\hat U}^{\s 0}_{\cal S}(\beta,t_0)\big|0\big\rangle_{\cal S}~\big|\beta,0\big\rangle_{\cal TD}\bigg]~+~\sum_{\alpha,t> t_0}^\beta~\chi(t)~\psi(\alpha|t)~\bigg[{\hat U}_{\cal S}(\beta,t)\big|\alpha\big\rangle_{\cal S}~\big|t,\alpha\big\rangle_{\cal TD}\bigg].
\end{eqnarray}
For future discussions, it is essential to keep in mind that $ | t, \alpha \rangle_{\cal TD} $ should not be recognized only as the physical state of $\cal TD$, but mainly by the information about $ \cal CS $ that this state stores. Thus, as $\beta>t$, we will refer to the state $|t,\alpha\rangle_{\cal TD}$ in the past as the memory of the event ``the system $\cal S$ in the state $|\alpha\rangle_{\cal S}$ with the clock $\cal C$ reading $t$.'' It is important to note that this particular interpretation is valid since the detector registers information about $\cal S$ by interacting directly with it, and the timer instantly recognizes the detection of $\cal S$. On this basis, Eq.~(\ref{Psi}) aims to describe $\cal CS$ from the perspective of $\cal TD$.

\subsection{Two causally connected events}
\label{form2e}

In this section, we will extend the calculation above to two events causally connected by assuming two consecutive measurements in the same system $ \cal S $. Here, causality means that the occurrence of the second event depends on the happening of the first event. In this regard, we consider two pairs of $ \cal TD $ so that, similarly to Eq.~(\ref{initial}), the initial condition is
\begin{eqnarray}\label{initialm}
\big|\psi(t_0)\big\rangle_{\cal R}=\big|0\big\rangle_{\cal S}~\big|t_0,0\big\rangle_{{\cal T}_1{\cal D}_1}~\big|t_0,0\big\rangle_{{\cal T}_2{\cal D}_2},
\end{eqnarray}
where ${\cal T}_1{\cal D}_1$ and ${\cal T}_2{\cal D}_2$ register the measurements $1$ and $2$ respectively. From now on, for the sake of understanding, let us consider the simplest case in which the operators defined in Eq.~(\ref{inf1}) satisfy ${\hat U}^{\s 0}_{\cal S}(t_{\s (j+1)},t_{\s (j)})\approx {\hat U}^{\s m}_{\cal S}(t_{\s (j+1)},t_{\s (j)}) \approx {\hat U}_{\cal S}(t_{\s (j+1)},t_{\s (j)})$. For two causally related events, the first step of the evolution~(\ref{Schr}) is the same as that of Eq.~(\ref{inf1}), but with the presence of the state $\big|t_{\s (1)},0\big\rangle_{{\cal T}_2{\cal D}_2}$ at the end of each branch of $|\psi(t_{\s (1)})\rangle_{\cal R}$. It should be warned not to confuse the labels $1$ and $2$ with $(1)$ and $(2)$: $j=1,2$ specify the events, while $(j)$ the possible values of $\beta$, $t_{ 1}$, and $t_{2}$ in the discrete evolution ($t_{\s (1)},t_{\s (2)},\hdots$). For simplicity, let us assume that the second measurement starts right after the end of the first one.

Under these circumstances, the second step of our schematic evolution is
\begin{eqnarray}\label{infN4m}
\big|\psi(t_{\s (2)})\big\rangle_{\cal R}&=&\sqrt{1-\delta p_{1\s(1)}}~{\rm e}^{\varphi_{ 1\s(1)}}~\sqrt{1-\delta p_{ 1\s(2)}}~{\rm e}^{\varphi_{2\s (2)}}~\bigg[{\hat U}_{\cal S}(t_{\s (2)},t_0)~\big|0\big\rangle_{\cal S}\bigg]~\big|t_{\s (2)},0\big\rangle_{{\cal T}_1{\cal D}_1}~\big|t_{\s (2)},0\big\rangle_{{\cal T}_2{\cal D}_2}
~\nonumber\\
&+&\sqrt{1-\delta p_{1\s (1)}}~{\rm e}^{\varphi_{1\s (1)}}~\sqrt{\delta p_{1\s (2)}}~\bigg[\sum_{\alpha_1}~{\hat M}_{\alpha_1}~{\hat U}_{\cal S}(t_{\s (2)},t_0)~\big|0\big\rangle_{\cal S}\bigg]~\big|t_{\s (2)},\alpha_{ 1}\big\rangle_{{\cal T}_1{\cal D}_1}~\big|t_{\s (2)},0\big\rangle_{{\cal T}_2{\cal D}_2}\nonumber\\
~\nonumber\\
&+&\sqrt{\delta p_{1\s (1)}}~\sqrt{1-\delta p_{2\s (2,1)}}~{\rm e}^{\varphi_{2\s (2)}}~\bigg[~{\hat U}_{\cal S}(t_{\s (2)},t_{\s (1)})~\sum_{\alpha_1}~{\hat M}_{\alpha_1}~{\hat U}_{\cal S}(t_{\s (1)},t_0)~\big|0\big\rangle_{\cal S}\bigg]~\big|t_{\s (1)},\alpha_{ 1}\big\rangle_{{\cal T}_1{\cal D}_1}~\big|t_{\s (2)},0\big\rangle_{{\cal T}_2{\cal D}_2}\nonumber\\
~\nonumber\\
&+&\sqrt{\delta p_{1\s (1)}}~\sqrt{\delta p_{2\s (2,1)}}~\bigg[\sum_{\alpha_2}~{\hat M}_{\alpha_2}~{\hat U}_{\cal S}(t_{\s (2)},t_{\s (1)})~\sum_{\alpha_1}{\hat M}_{\alpha_1}~{\hat U}_{\cal S}(t{\s (1)},t_0)~\big|0\big\rangle_{\cal S}\bigg]~\big|t_{\s (1)},\alpha_{ 1}\big\rangle_{{\cal T}_1{\cal D}_1}~\big|t_{\s (2)},\alpha_{ 2}\big\rangle_{{\cal T}_2{\cal D}_2},\nonumber\\
\end{eqnarray}
Here, the first two branches come from the evolution of the unmeasured contribution of $|\psi(t_{\s(1)})\rangle_{\cal R}$. The first term of $|\psi(t_{\s (2)})\rangle_{\cal R}$ illustrates the situation where no events happen during the clock reading interval $[t_0,t_{\s (2)}]$. In the second one, the first event takes place at the second step when the clock reads $t=t_{\s (2)}$. The last two contributions in Eq.~(\ref{infN4m}) result from the evolution of the branch of $ | \psi (t_{\s (1)}) \rangle_{\cal R} $ in which the first event took place at $t_{\s (1)}$. In the third (fourth) expression of~(\ref{infN4m}), the second measurement does not happen (happens) at $t_{\s (2)}$, and $\delta p_{2\s(2,1)}$ is the probability of the second event occurring with the clock reading $t_{\s(2)}$ regardless of the outcome $\alpha_2$, given that the first measurement took place at $t_{(1)}$. Keep doing the evolution until $\beta=t_{\s (N)}$, we obtain a similar expression to~(\ref{Psi}) given by
\begin{eqnarray}\label{infNN}
\big|\psi\big(\beta\big)\big\rangle_{\cal R}&=&\Gamma_{1}(\beta,t_0)~\bigg[{\hat U}_{\cal S}(\beta,t_0)~\big|0\rangle_{\cal S}~\big|\beta,0\big\rangle_{{\cal T}_1{\cal D}_1}~\big|\beta,0\big\rangle_{{\cal T}_2{\cal D}_2}\bigg]\nonumber\\
~\nonumber\\
&+&\sum_{\substack{\alpha_1 \\ t_1> t_0}}^\beta~\Gamma_{2}(\beta,t_{ 1})~\psi_{\cal S}(\alpha_{ 1}|t_{ 1})~\chi_{1}(t_{ 1})~\bigg[{\hat U}_{\cal S}(\beta,t_{1})~\big|\alpha_{ 1}\big\rangle_{\cal S}~\big|t_{ 1},\alpha_{ 1}\big\rangle_{{\cal T}_1{\cal D}_1}~\big|\beta,0\big\rangle_{{\cal T}_2{\cal D}_2}\bigg]
~\nonumber\\
&+&\sum_{\substack{\alpha_2 \\ t_2> t_1}}^{\beta}~\sum_{\substack{\alpha_1 \\ t_1> t_0}}^{t_1<\beta}~~\psi_{\cal S}(\alpha_{2}|t_{ 2};\alpha_{1},t_{ 1})~\chi_{2}(t_{ 2}|t_{ 1})~\psi_{\cal S}(\alpha_{ 1}|t_{ 1})~\chi_{ 1}(t_{ 1})~\bigg[{\hat U}_{\cal S}(\beta,t_{ 2})\big|\alpha_{\s 2}\big\rangle_{S}~\big|t_{ 1},\alpha_{1}\big\rangle_{{\cal T}_1{\cal D}_1}~\big|t_{ 2},\alpha_{ 2}\big\rangle_{{\cal T}_2{\cal D}_2}\bigg],\nonumber\\
\end{eqnarray}
where
\begin{eqnarray}\label{coef2}
\psi_{\cal S}(\alpha_{ 2}|t_{ 2};\alpha_{ 1},t_{ 1})={_{\cal S}\big\langle}\alpha_2\big|{\hat U}_{\cal S}(t_{2},t_{ 1})\big|\alpha_{ 1}\big\rangle_{\cal S}
\end{eqnarray}
is the Schr\"odinger wave function of $\cal S$ at the instant $t_{ 2}$ in the case where the first event $(t_{ 1},\alpha_{ 1})$ took place. Hence, $|\psi_{\cal S}(\alpha_{ 2}|t_{ 2};\alpha_{ 1},t_{ 1})|^2$ is the probability of the system having been measured in the state $|\alpha_{ 2}\rangle_{\cal S}$ given that the clock read $t_{2}$, and that $\cal S$ was measured in $|\alpha_{ 1}\rangle_{\cal S}$ with $\cal C$ reading $t_{ 1}$. Finally, the time probability amplitudes of the first and second events are 
\begin{eqnarray}\label{dP2}
\chi_{ 1}(t_{\s (k)})=\sqrt{\delta p_{ 1\s(k)}}~\prod_{q=1}^{k-1}\sqrt{1-\delta p_{1\s (q)}}~{\rm e}^{i\varphi_{1\s (q)}}~~~~{\rm and}~~~~\chi_{ 2}(t_{\s (\ell)}|t_{\s (k)})=\sqrt{\delta p_{\s (\ell,k)}}\prod_{r=k+1}^{\ell-1}\sqrt{1-\delta p_{2\s (r,k)}}~{\rm e}^{i\varphi_{2\s (r,k)}}.
\end{eqnarray}

Let us carefully analyse Eqs.~(\ref{infNN})-(\ref{dP2}). In the first expression of Eq.~(\ref{infNN}), no events happen between $[t_0,\beta]$, thus both ${{\cal T}_{1}{\cal D}_{ 1}}$ and ${{\cal T}_{2}{\cal D}_{ 2}}$ have no information about $\cal S$. In the second term of Eq.~(\ref{infNN}), $|\alpha_{ 1},t_{ 1}\rangle_{{\cal D}_1{\cal T}_1}~|0,\beta\rangle_{{\cal D}_2{\cal T}_2}$ represents the situation in which the system $\cal S$ was in the state $|\alpha_{ 1}\rangle_{\cal S}$ with the clock reading $t_{ 1}$, and no event happens from $t_{ 1}$ to $\beta$. Besides, $|\chi_{ 1}(t_{ (k)})|^2$ is the probability of the first event having happened with the clock reading $t_{\s (k)}$, regardless of the outcome $\alpha_{ 1}$. Finally, the last line of $|\psi(\beta)\rangle_{\cal R}$ describes the case where the two events occurred within the interval $]t_0,\beta]$. Thus, $|\alpha_{ 1},t_{ 1}\rangle_{{\cal D}_1{\cal T}_1}~|\alpha_{ 2},t_{ 2}\rangle_{{\cal D}_2{\cal T}_2}$ means ``$\cal S$ was in the state $|\alpha_{ 1}\rangle_{\cal S}$ with the clock reading $t_{ 1}$, and in the state $|\alpha_{ 2}\rangle_{\cal S}$ at $t_{ 2}$.'' Also, $|\chi_{ 2}(t_{\s (\ell)}|t_{\s (k)})|^2$ is the probability of the second measurement having taken place at time $t_{\s (\ell)}$, given that the first measurement happened at time $t_{\s (k)}<t_{\s (\ell)}$, independent of the outcome. This causal dependence can be verified by observing the sum over $t_{ 2}$ in Eq.~(\ref{infNN}), in which $t_{ 2}>t_{ 1}$, as well as the product operator on the right side of Eq~(\ref{dP2}), which begins at $k+1$. It ensures that the second event takes place only after the first one. The joint probability of these two events is the modulus squared of the wave function outside the brackets of the last sum of Eq.~(\ref{infNN}),
\begin{eqnarray}\label{probx}
P(\alpha_{ 1},t_{1};\alpha_{ 2},t_{ 2})=\big|\psi_{\cal  S}(\alpha_{ 2}|t_{ 2};\alpha_{ 1},t_{ 1})\big|^2~\big|\chi_{ 2}(t_{ 2}|t_{ 1})\big|^2 ~\big|\psi_{\cal S}(\alpha_{1}|t_{ 1})\big|^2\big|\chi_{ 1}(t_{ 1})\big|^2.
\end{eqnarray}

Finally, it should be noticed that the conditioned state~(\ref{infNN}) has an indefinite number of events, so that if an experimentalist decides to check the records of $\cal TD$ at this moment, she can observe none, $1$, or $2$ events. Although Eq.~(\ref{infNN}) can predict observations of an external experimentalist, as discussed earlier, a rigorous analysis would demand the conditioning of $||\Psi\rangle$ on the state of the experimentalist's brain and not on $|\beta\rangle_{\cal C}$. Thus, as $\cal C$ is by definition isolated, in Sec.~\ref{qme}, we will describe events through $||\Psi\rangle$, and not $|\psi(\beta)\rangle_{\cal R}$. But first, let us investigate how to deal with events in more elementary circumstances, i.e., without resorting to an artificial laboratory timer to record the detection time.

\subsection{The environment as a timer}
\label{env}

Aiming a more fundamental characterization of quantum events, we will discuss the possibility of a detector with many degrees of freedom (more specifically, the environment) developing the role of a set of timers $\cal T$. Here, we will explicitly refer to the environment by redefining
$\cal D$ of the last section as $\cal DE$. Let us start with a single detector being continuously monitored by the environment around it~(see, for instance, Ref.~\cite{Hal2}). Under this interaction, any macroscopic change in the detector's state is ``instantaneously'' recognized by the environment through a measurement-like process similar to Eq.~(\ref{VN}). It means that the time scale of the interaction between the system and detector [$\Delta t_{\cal SD} $ in Eq.~(\ref {VN})] is significantly longer than the time $\delta t_{\cal DE} $ required for the environment to read information about $ \cal S $ stored in the detector. This assumption is the Markov approximation commonly used to obtain, for instance, the Lindblad equation for open quantum systems~\cite{Zeh}.

Under these circumstances, at every interval $ \delta t_{\cal DE} $, a subsystem of the environment acquires information about the detector and then rapidly departs from it, not interacting with the detector again. This interaction defines the pointer states of the detector~\cite{Zu}. As $ \delta t_{\cal DE} $ is the smallest time scale of the problem, it is convenient to assume that the steps $ \delta t $ of the evolution~(\ref{Schr}) is approximately $ \delta t_{\cal DE}$. Thus, at every $\delta t$, the detector interacts with a new degree of freedom of the environment in such a way that the detector always faces the environment in its initial state. This kind of interaction, but in a completely different context, can be found in Ref.~\cite{Zeh}.

Let us consider the evolution for the case of a single event of Sec.~\ref{form1e}, with the environment interacting with $\cal D$ and in the absence of $\cal T$. In this context, the initial state~(\ref{initial}) becomes
\begin{eqnarray}\label{initiale}
\big|\psi(t_0)\big\rangle_{\cal R}&=&\big|0\big\rangle_{\cal S}~\big|0\big\rangle_{\cal D}~\big|r_{\s (1)}, r_{\s (2)}, r_{\s (3)}, \hdots\big\rangle_{\cal E},
\end{eqnarray}
where $|r_{\s (1)}\rangle$, $|r_{\s (2)}\rangle$, $\hdots$ are the ready states of the subsystems of the environment that eventually acquire information about $\cal S$ via interaction with $\cal D$. As  in the circumstances of this section, the evolution of $\cal S$ and $\cal D$ is the same as the one in Sec.~\ref{form1e}, let us exclusively focus on the dynamic of the environment. In the first step of the evolution~(\ref{inf1}), the environment changes according to
\begin{eqnarray}\label{trans1}
\big|r_{\s (1)}, r_{\s (2)}, r_{\s (3)}, \hdots\big\rangle_{\cal E}~\rightarrow ~ \big|0_{\s (1)}, r_{\s (2)},r_{\s (3)}, \hdots \big\rangle_{\cal E}~;~\big| \alpha_{\s (1)}, r_{\s (2)}, r_{\s (3)}, \hdots \big \rangle_{\cal E},
\end{eqnarray}
where at $t_{\s (1)}$ becomes a superposition of the two last states. In the branch of $|\psi(t_{\s (1)})\rangle_{\cal R}$ where there is no detection, the environment records the state $|0\rangle_{\cal D}$ of the detector, and thus the state of the environment is $ |0_{\s (1)}, r_{\s (2)},r_{\s (3)}, \hdots \rangle_{\cal E}$. In the other branch of Eq.~(\ref{inf1}), the detector measures the system and instantly transmits this information to the environment's subsystem $ (1) $. As a result, the state of the environment is $ | \alpha_{\s (1)}, r_{\s (2)}, r_{\s (3)}, \hdots \rangle_{\cal E} $. In both branches of Eq.~(\ref{inf1}), the dispersion of the subsystem $ (1) $ makes the detector interact with the environment's degree of freedom $ (2) $ at $t_{\s (2)}$.

From $t_{\s (1)}$ to $t_{\s (2)}$ [see Eq.~(\ref{psi2})], the environment makes the transitions
\begin{eqnarray}\label{trans2}
\big|0_{\s (1)}, r_{\s (2)}, r_{\s (3)}, \hdots\big\rangle_{\cal E}~&\rightarrow& ~\big|0_{\s (1)}, 0_{\s (2)},r_{\s (3)}, \hdots \big\rangle_{\cal E}~;~\big| 0_{\s (1)}, \alpha_{\s (2)}, r_{\s (3)}, \hdots \big \rangle_{\cal E},~~{\rm and}\nonumber\\~\nonumber\\
\big|\alpha_{\s (1)}, r_{\s (2)}, r_{\s (3)}, \hdots\big\rangle_{\cal E}~&\rightarrow& ~\big| \alpha_{\s (1)}, \alpha_{\s (2)}, r_{\s (3)}, \hdots \big \rangle_{\cal E}.
\end{eqnarray}
Here, the branch $| 0_{ \s (1)}, r_{\s (2)}, r_{\s (3)}, \hdots \rangle_{\cal E}$ splits into two parts. In one of them, the detector remains without measuring the system, and hence the environment ``visualizes'' the detector again in its initial state, becoming $ |0_{\s (1)}, 0_{\s (2)}, r_{\s (3)}, \hdots \rangle_{\cal E} $. In the other branch of $ | \psi (t_{\s (2)}) \rangle_{\cal R} $ coming from $ | 0_{\s (1)}, r_{\s (2)}, r_{\s (3)}, \hdots \rangle_{\cal E} $, the detector measures the system, and then the environment's state becomes $ | 0_{\s (1)}, \alpha_{\s (2)}, r_{\s (3)} \hdots \rangle_{ \cal E} $. Note that in Eq.~(\ref{trans2}), the information about $\cal S$ is stored in the subsystem (2) of the environment, whereas the subsystem (1), which is away from the detector at $t_{\s (2)}$, remains in the state $ |0_{\s (1)} \rangle_{\cal E} $. Regarding the branch $| \alpha _ {\s (1)}, r_{\s (2)}, r_{\s (3)}, \hdots \rangle_{\cal E}$ at $ t_{\s (1)}$, as we are dealing with a single event, the environment changes to $ | \alpha_{\s (1)} , \alpha_{\s (2)}, r_{\s (3)}, \hdots \rangle _ {\cal E} $ at $ t_{\s (2)}$. The information about $ \cal S $ recorded in the detector $\cal D$ is now also stored in the environment's subsystem $ (2) $.

Keeping this evolution until the step $ N $, in the branch of $ | \psi (t_{\s (N)}) \rangle_{\cal R} $ where the measurement happened at the instant $ t_{\s (k)} $, the environment has state
\begin{eqnarray}\label{infNe}
\big|t_{\s (k)},\alpha\big\rangle_{\cal E}\equiv\big|0_{\s (1)},0_{\s (2)},\hdots,0_{\s (k-1)},\alpha_{\s (k)},\hdots,\alpha_{\s (N)},r_{\s (N+1)},r_{\s (N+2)},\hdots\big\rangle_{\cal E}.
\end{eqnarray}
It is readily verified that the position $ (k) $ of the first subsystem of $\cal E$ that registers $ \alpha $ indicates that the clock $\cal C$  read $ t = t_{\s (k)} $ at the moment of the event.  Therefore, for different clock readings $t_{\s (k)} $ at the instant of the event, $\cal E$ records the state of the system in distinct arrangements. As the states $ \{|0_{\s (1)},0_{\s (2)},\hdots,\alpha_{\s (k)},\hdots,\alpha_{\s (N)},r_{\s (N+1)},\hdots \rangle_{\cal E} \}_{k=1,2,\hdots,N}$ are orthogonal to each other, they can unambiguously register the possible events of the set $\{ (t_{\s (k)}, \alpha)\}_{k=1,2,\hdots,N}$. In other words, by measuring the environment subsystems, it is possible, in principle, to find out the state $|\alpha\rangle_{\cal S}$ of $\cal S$ and the clock reading when $\cal S$ was in this state. For instance, if $\alpha=\pm$ and the system was in $|+\rangle_{\cal S}$ $\big(|-\rangle_{\cal S}\big)$ at $t_{\s (2)}$ $\big(t_{\s (3)}\big)$, the state of the environment at $t=t_{\s (N)}$ is
\begin{eqnarray}\label{infNex}
\big|t_{\s (2)},+\big\rangle_{\cal E}\equiv\big|0_{\s(1)},+_{\s (2)},+_{\s (3)},\hdots,+_{\s (N)},r_{\s (N+1)},\hdots\big\rangle_{\cal E}~~ \bigg(~\big|t_{\s (3)},-\big\rangle_{\cal E}\equiv\big|0_{\s (1)},0_{\s (2)},-_{\s (3)},\hdots,-_{\s (N)},0_{\s (N+1)},\hdots \big\rangle_{\cal E}~\bigg).
\end{eqnarray}
Notice that we could consider the environment interacting directly with $ \cal S $, and thereby also disregard the detector in the measurement process and still obtain the same state for $\cal R$ as the one with $\cal TD$. For example, in the case where a particle passes through a cloud chamber~\cite{Mott}, we could use, in principle, the formalism presented here to describe this particle from the perspective of the supersaturated vapor of water inside the chamber.

Therefore, when the information of the moment of the event is recorded in some degree of freedom of $\cal E$, the conditioned state of $\cal R$ should be given by equations such as~(\ref{Psi}) and~(\ref{infNN}), but with $\cal TD$ being replaced by $\cal DE$. Nevertheless, because usually we do not have access to the environment's degree of freedom, we do not detect $P(t,\alpha)$, but $P(\alpha)$ as written in Eq.~(\ref{prob2}). Besides, notice that if $\Delta t_{\cal SD}$ is short enough, Eq.~(\ref{infN2}) with $\cal E$ substituting $\cal T$ becomes
\begin{eqnarray}\label{Psia}
\big|\psi(t_{\s (N)})\big\rangle_{\cal R}\approx \sum_\alpha~\psi_{\cal S}(\alpha|t_0)~\big|\alpha\big\rangle_{\cal S}~\big|\alpha \big\rangle_{\cal D}~\Bigg[\sum_{k=1}^{{N}}~\chi(t_{\s k})~\big|t_{\s (k)},\alpha\big\rangle_{\cal E}\Bigg].
\end{eqnarray}
In this simpler scenario, the inaccessibility of the information about $\cal S$ stored in $\cal E$ allows us to conveniently define $|\alpha\rangle_{\cal E}\equiv \sum_{k=1}^{{\s N}}~\chi(t_{\s k})~|t_{\s (k)},\alpha\rangle_{\cal E}$ so that
\begin{eqnarray}\label{tr}
\big|\psi(t_{\s (N)})\big\rangle_{\cal R}=\sum_\alpha~\psi_{\cal S}(\alpha|t_0)~\big|\alpha\big\rangle_{\cal S}~\big|\alpha\big\rangle_{\cal D}~\big|\alpha\big\rangle_{\cal E}
\end{eqnarray}
becomes the measurement~(\ref{VN}) taking into account the environment-induced decoherence. In this case,  $P(\alpha)\approx |\psi_{\cal S} (\alpha | t_{0 })|^2 $, as expected.

Finally, notice that the generalization of the environment's state for an arbitrary number of events is straightforward. For instance, in the case of $2$ events that happened at the instants $ t_{\s (k)} $ and $ t_{\s (\ell)} $, the state of the environment for $ \beta = t_{\s (N)}$ is
\begin{eqnarray}\label{infNe2}
\big|t_{\s (k)},\alpha_{\s 1}\big\rangle_{\cal E}~\big|t_{\s (\ell)},\alpha_{\s 2}\big\rangle_{\cal E}\equiv\big|0_{\s (1)},\hdots,\alpha_{\s 1\s (k)},\hdots,\alpha_{\s 1\s (\ell)},\hdots,\alpha_{\s 1\s (N)},r_{\s (N+1)},\hdots\big\rangle_{{\cal E}_1}~\big|0_{\s (1)},\hdots,0_{\s (k)},\hdots,\alpha_{\s 2\s (\ell)},\hdots,\alpha_{\s 2\s (N)},r_{\s (N+1)},\hdots\big\rangle_{{\cal E}_2}.\nonumber\\
\end{eqnarray}
Here, $ {\cal E}_{\s 1} $ and $ {\cal E}_{\s 2} $ represent the environment around the detector ${\cal D}_{\s 1} $ and $ {\cal D}_{\s 2} $ respectively. After this discussion about events via the conditioned state $|\psi(\beta)\rangle_{\cal R}$ (with either timers and environment) let us now formalize a more general description in terms of $||\Psi\rangle$.

\section{Events and the emergence of time from the perspective of $||\Psi\rangle$}
\label{qme}

\subsection{A single event}
\label{qme1}

In the previous section, although we obtained a joint probability distribution for events as proposed in the introduction, we still resorted to a relative approach between $ \cal R $ and $ \cal C $, $ |\psi(\beta)\rangle_{\cal R}$. Remember that the presence of an external observer to the formalism allowed us to make the conditioning $ |\psi(\beta)\rangle_{\cal R}={_{\cal C} \langle} \beta || \Psi \rangle $. In contrast, the focus now is exclusively on describing $ \cal CS $ from the perspective of any observer $ \cal TD $, and hence, as $\cal TD$ do not interact with the clock $\cal C$, we will not condition $||\Psi\rangle$ on a specific state $|\beta\rangle_{\cal C}$. Thus, if an experimentalist has to be taken into account, her observation must be included in the formalism as an event. Besides, it should be remembered that, without addressing the nature of time, we will remain calling $\beta$ and $t$ as measures of time until Sec.~\ref{time}. Let us begin by calculating $||\Psi\rangle$ for a single event. By substituting Eq.~(\ref{Psi}) into Eq.~(\ref{PWsol}) with $\cal R=SDT$, we obtain
\begin{eqnarray}\label{PsiG}
\big|\big|\Psi\big\rangle=\sum_{\beta \geq t_0}~~\Gamma(\beta,t_0)~\bigg[\big|\beta\big\rangle_{\cal C}~{\hat U}_{\cal S}(\beta,t_0)\big|0\big\rangle_{\cal S}~\big|\beta,0\big\rangle_{\cal TD}\bigg]~+\sum_{\beta\geq t_0}~~\sum_{\substack{\alpha \\ t> t_0}}^{\beta}~~\chi(t)~\psi_{\cal S}(\alpha|t)~\bigg[\big|\beta\big\rangle_{\cal C}~{\hat U}_{\cal S}(\beta,t)\big|\alpha\big\rangle_{\cal S} ~\big|t,\alpha\big\rangle_{\cal TD}\bigg].\nonumber\\
\end{eqnarray}
Here, $||\Psi\rangle$ takes into account both the occurrence and non-occurrence of the event in the interval of the clock reading $ [t_0, \beta] $, for $\beta=t_0,t_{\s (1)},t_{\s (2)},\hdots$ . Now, we will carefully investigate Eq.~(\ref{PsiG}) to obtain a suitable notation for describing events.

A relevant feature of $||\Psi\rangle$ can be verified by rewriting the second term of Eq.~(\ref{PsiG}) using the relation $\sum_{\beta \geq t_0}\sum_{t>t_0}^\beta=\sum_{t>t_0}\sum_{\beta \geq t}$. By highlighting only this term and isolating the state for $\beta=t$ from the sum over $\beta$, Eq.~(\ref{PsiG}) becomes
\begin{eqnarray}\label{PsiGft}
\big|\big|\Psi\big\rangle &=&\hdots~+\sum_{\substack{\alpha \\ t> t_0}}~\chi(t)~\psi_{\cal S}(\alpha|t)~\big|t\big\rangle_{\cal C}\big |\alpha\big\rangle_{\cal S}~\big|t,\alpha\big \rangle_{\cal TD}+\sum_{\substack{\alpha \\ t> t_0}}~~\sum_{\beta >t}~~\chi(t)~\psi_{\cal S}(\alpha|t)~\big|\beta\big\rangle_{\cal C}~{\hat U}_{\cal S}(\beta,t)\big|\alpha\big\rangle_{\cal S}~\big|t,\alpha\big \rangle_{\cal TD}.
\end{eqnarray}
By analysing the first contribution of Eq.~(\ref{PsiGft}), we observe the correlation between the state of $\cal CS$ and the information acquired by $\cal TD$. Let us call this $ \alpha $-time agreement as the quantum event $ (t, \alpha) $ whose state is defined as
\begin{eqnarray}\label{event}
\big|\big|t,\alpha\big\rangle \equiv \big|t,\alpha\big\rangle_{\cal CS}~\big|t,\alpha\big \rangle_{\cal TD}.
\end{eqnarray}
As discussed earlier, since the modulus squared of $ \psi (\alpha | t) \chi (t) $ [coefficient of the first term of Eq.~(\ref{PsiGft})] is the probability of $ (t, \alpha) $, we will call this amplitude the wave function $\Psi (t, \alpha)$ of this event. Here, we use the two bars to indicate that this state contains the pairs $ \cal TD $ and $ \cal CS $ recording the same information.

It should be remembered that in the previous sections, because $\beta>t$, we referred to $|t,\alpha\rangle_{\cal TD}$ in $|\psi(\beta)\rangle_{\cal R}$ as an event that occurred in the past. In contrast, because of the agreement ($\beta=t$) between $\cal C$ and $\cal T$, $||t,\alpha\rangle$  represents $\cal CS$ at the moment of the observation of $\cal TD$ (the event itself), and hence $||t,\alpha\rangle$ will be referred in the present as ``from the perspective of $ \cal TD $, the system $ \cal S $ has the property $\alpha$ (is in the state $ |\alpha\rangle) $ and the clock $\cal C$ reads $t$.''  Thus, $|\Psi (t, \alpha)|^2$ is the probability of the system being in the state $|\alpha\rangle_{\cal S}$ and the clock reading $t$ from $\cal TD$'s point of view. As remarked before, unlike the classical treatment for events, Eq.~(\ref{event}) shows that a quantum event happens with the exchange of information between physical systems, so that the event can be seen as the very recording of this information. From this standpoint, those events studied in special relativity, such as lightning bolts striking a train, should be extended to the QM domain by taking into account, for instance, a state depicting an agreement similar to $||t,\alpha\rangle$ between the lightning bolt and the train.

Now, by inspecting the second contribution of Eq.~(\ref{PsiGft}), we verify that in the product state $|\beta\rangle_{\cal C}{\hat U}_{\cal S}(\beta,t)\big|\alpha\big\rangle_{\cal S}|t,\alpha\rangle_{\cal TD}$, the reading of $\cal C$ is evaluated after the occurrence of the event $(t,\alpha)$, i.e., $\beta>t$. Thus, similarly to the last section, the state $ | t, \alpha \rangle_{\cal TD} $ is a memory of the event $(t,\alpha)$ that occurred with the clock reading $ \beta=t $. Notice that there is a sum over $\beta>t$ in Eq.~(\ref{PsiGft}) because, in our model, the event remains recorded in the detector and timer after its occurrence. Finally, let us turn our attention to the first term of Eq.~(\ref{PsiG}), which concerns the physical scenario before the occurrence of the event $(t,\alpha)$. In this branch, the modulus squared of $ \Gamma(\beta, t_0) $ provides the probability for the event not to happen in the interval $ [t_0, \beta] $. Thus, consider $ \Psi(t, 0) \equiv \Gamma(t, t_0) $ and, in the spirit of Eq.~(\ref{event}), define
\begin{eqnarray}\label{nevent}
\big|\big|\beta,0\big\rangle \equiv \big|\beta\big\rangle_{\cal C}~{\hat U}_{\cal S}(\beta,t)\big|0\big\rangle_{\cal S}~\big|\beta,0\big\rangle_{\cal TD}
\end{eqnarray}
as the state associated with the absence of events. In Eq.~(\ref{nevent}), the clock and the timer are synchronized, but $ \cal D $ has no information about $ \cal S $. Also, notice that $|t,\alpha\big\rangle$ and $|\beta,0\rangle$ are orthogonal to each other.

By using the states (\ref{event}) and (\ref{nevent}), we will rewrite $||\Psi\rangle$ by joining the first term of Eq.~(\ref{PsiG}) and the first one of Eq.~(\ref{PsiGft}) --- which refer to the non-occurrence and occurrence of the event respectively --- in the same sum. To this end, first, let us make the convenient change of variables: $\beta \rightarrow t_{ 1}$ ($\beta \rightarrow t_{2}$) in the first (second) term of Eq.~(\ref{PsiG}), and  $t \rightarrow t_{ 1}$ in Eq.~(\ref{PsiGft}). Besides, consider $\alpha\rightarrow \alpha_{ 1}$. Now, by using Eqs.~(\ref{event}) and (\ref{nevent}), and the first terms of Eqs.~(\ref{PsiG}) and~(\ref{PsiGft}) expressed in a sum over $t_{ (1)}$, $ || \Psi \rangle $ of Eq.~(\ref{PsiG}) becomes
\begin{eqnarray}\label{PsiGG}
\big|\big|\Psi\big\rangle&=&\sum_{\substack{a_1 \\ t_1\geq t_0}}~\Psi \big(t_{ 1},a_{1}\big)~\big|\big|t_{ 1},a_{1}\big\rangle~+~\sum_{t_{ 2}> t_{1}}~\sum_{\substack{\alpha_1 \\ t_1> t_0}}~\chi(t_{ 1})~\psi(\alpha_{ 1}|t_{ 1})~\bigg[\big|t_{ 2}\big\rangle_{\cal C}~{\hat U}_{\cal S}(t_{ 2},t_{ 1})\big|\alpha_{ 1}\big\rangle_{\cal S} ~\big|t_{ 1},\alpha_{1}\big\rangle_{\cal TD}\bigg],
\end{eqnarray}
with $a_{ 1}=0$ and $a_{ 1}=\alpha_{ 1}$ describing the non-occurrence and occurrence of the event respectively.

Lastly, the second contribution of Eq.~(\ref{PsiGG}), which is the branch associated with memory, should be written in the same notation as the first one. To this end, consider a collection of detectors and timers such that ${\cal TD}={\cal T}_{ 1}{\cal D}_{1}~{\cal T}_{ 2}{\cal D}_{ 2}~{\cal T}_{ 3}{\cal D}_{ 3}\hdots$. Now, let us consider this collection as being the observer. Because we are still dealing with a single event, these additional detectors do not measure $ \cal S $. By assuming that each timer has the same initial condition as $ {\cal T}_{1} $ and is synchronized with $ \cal C $, the states~(\ref{event}) and~(\ref{nevent}) in the presence of the additional detectors and timers become $||t_{ 1},\alpha_{ 1}\rangle \equiv |t_{ 1},\alpha_{ 1}\rangle_{\cal CS}~|t_{ 1},\alpha_{ 1} \rangle_{{\cal T}_1{\cal D}_1}~|t_{ 1},0\rangle_{{\cal T}_2{\cal D}_2}~|t_{ 1},0\rangle_{{\cal T}_3{\cal D}_3}~\hdots$ and $||t_{ 1},0\rangle \equiv |t_{ 1}\rangle_{\cal C}~{\hat U}_{\cal S}(t_{1},t_{ 0})|\alpha_{ 1}\rangle_{\cal S}~|t_{ 1},0\rangle_{{\cal T}_1{\cal D}_1}~|t_{ 1},0\rangle_{{\cal T}_2{\cal D}_2}~|t_{ 1},0\rangle_{{\cal T}_3{\cal D}_3}~\hdots$, respectively. Notice that in both cases, the new detectors have no information about $ \cal S $. Also, despite the presence of numerous ${\cal T}_j{\cal D}_j$, as our focus is events, the state of the event $(t_{1},\alpha_{1})$ is represented by the single ket $||t_{ 1},\alpha_{ 1}\rangle$. Under these circumstances, by following the same notation as the first term of Eq.~(\ref{PsiGG}), the memory state [ket within the brackets in Eq.~(\ref{PsiGG})] is rewritten as
\begin{eqnarray}
\label{mevent}
\big|\big|t_{ 1},\alpha_{ 1};t_{ 2},0\big\rangle \equiv \big|t_{ 2}\rangle_{\cal C}~{\hat U}_{\cal S}(t_{ 2},t_{ 1})\big|\alpha_{ 1}\big\rangle_{\cal S}~\big|t_{ 1},\alpha_{ 1}\big\rangle_{{\cal T}_1{\cal D}_1}~\big|t_{ 2},0\big\rangle_{{\cal T}_2{\cal D}_2}~\big|t_{ 2},0\big\rangle_{{\cal T}_3{\cal D}_3}~\hdots,
\end{eqnarray}
where ${\cal T}_{ 1}{\cal D}_{ 1}$ records the event $(t_{ 1},\alpha_{ 1})$ and ${\cal T}_{ 2}{\cal D}_{2}~{\cal T}_{3}{\cal D}_{ 3}\hdots$ do not have information about $\cal S$ with the clock reading $t_{2}$. Therefore, with the additional timers and detectors, $ || t_{ 1}, \alpha_{ 1}; t_{ 2}, 0 \rangle $ can be seen as the state related to the absence of a second event in the interval $]t_{ 1}, t_{2}]$, and the memory of the first event $(t_{ 1}, \alpha_{1 }) $. On the other hand, the state related to the occurrence of the second event, which must have zero amplitude since we are still dealing with a single event, can be written as
\begin{eqnarray}\label{mevent1}
\big|\big|t_{1},\alpha_{1};t_{ 2},\alpha_{ 2}\big\rangle \equiv \big|t_{1},\alpha_{ 1}\big\rangle_{\cal CS}~\big|t_{ 1},\alpha_{1}\big\rangle_{{\cal T}_1{\cal D}_1}~\big|t_{ 2},\alpha_{ 2}\big\rangle_{{\cal T}_2{\cal D}_2}~\big|t_{2},0\big\rangle_{{\cal T}_3{\cal D}_3}~\hdots,
\end{eqnarray}
with the $\alpha $-time agreement happening between $\cal CS $ and $ {\cal T}_{2} {\cal D}_{2} $, and including the memory of the first event $(t_{1}, \alpha_{ 1}) $.

Finally, according to Eq.~(\ref{PsiGG}), the wave functions of the states~(\ref{mevent}) and~(\ref{mevent1}) are $\Psi (t_{1}, \alpha_{1}; t_{2}, 0) = \Psi (t_{ 1}, \alpha_{1}) $ and $ \Psi (t_{1}, \alpha_{ 1}; t_{ 2}, \alpha_2) = 0 $ respectively. Then, by taking into account ${\cal T}_{2}{\cal D}_{ 2}~{\cal T}_{ 3}{\cal D}_{ 3}\hdots$, the global state $ || \Psi \rangle $ of Eq.~(\ref{PsiGG}) can be rewritten as
\begin{eqnarray}\label{PsiGGf}
\big|\big|\Psi\big\rangle=\sum_{\substack{a_1 \\ t_1\geq t_0}}~\Psi \big(t_{ 1},a_{ 1}\big)~\big|\big|t_{ 1},a_{ 1}\big\rangle~+~\sum_{\substack{a_2 \\ t_2> t_1}}~\sum_{\substack{\alpha_1 \\ t_1> t_0}}~\Psi \big(t_{ 1},\alpha_{ 1};t_{ 2},a_{ 2}\big) ~\big|\big|t_{ 1},\alpha_{ 1}; t_{ 2},a_{ 2}\big\rangle,
\end{eqnarray}
where $a_{ 2}=0$ [the memory contribution of Eq.~(\ref{PsiGG})] can now be seen as the absence of the second event, and $a_{ 2}=\alpha_{ 2}$ the occurrence of the second event. Here, $  \Psi (t_{ 1}, \alpha_{ 1}; t_{ 2}, 0)  $ is the probability amplitude of the second event not happening in the interval $] t_{1}, t_{ 2}] $ and the first event $(t_{ 1}, \alpha_{ 1}) $ having occurred. Notice that by Bayes' rule, this distribution can be written as $ | \Psi (t_{ 1}, \alpha_{ 1}; t_{ 2}, 0) |^2 = P (t_{ 2}, 0 | t_ { 1}, \alpha_{1})~| \Psi(t_{ 1}, \alpha_{ 1})|^2 $, with $P (t_{ 2}, 0 | t_ { 1}, \alpha_{ 1})=1$. Hence, as expected, since we are considering a single event, the probability of the second event not occurring is always 1, regardless of the value of $ t_{ 2} $. It should be noted that as the event states are orthogonal to each other, we have
\begin{eqnarray}\label{coef}
\big \langle t_{ 1}^{},a_{ 1}^{}\big|\big|t_{ 1}^{\prime},a_{ 1}^{\prime} \big\rangle=\delta_{t_{ 1}^{},t_{ 1}^{\prime}}~\delta_{a_{ 1}^{},a_{ 1}^{\prime}};~~\big \langle t_{ 1}^{},\alpha_{ 1}^{};t_{\s 2}^{},a_{ 2}^{}\big|\big|t_{ 1}^{\prime},\alpha_{ 1}^{\prime};t_{ 2}^{\prime},a_{ 2}^{\prime} \big\rangle=\delta_{t_{ 1}^{},t_{ 1}^{\prime}}~\delta_{\alpha_{ 1}^{},\alpha_{1}^{\prime}}~\delta_{t_{ 2}^{},t_{ 2}^{\prime}}~\delta_{a_{ 2}^{},a_{ 2}^{\prime}};~~{\rm and}~~\big \langle t_{ 1},a_{ 1}\big|\big|t_{ 1},\alpha_{1};t_{ 2},a_{ 2}\big\rangle=0,\nonumber\\
\end{eqnarray}
which imply that $\big \langle a_{ 1},t_{ 1}\big|\big|\Psi\big\rangle=\Psi(a_{ 1},t_{ 1})~~~{\rm and}~~~\big\langle \alpha_{ 1},t_{1};a_{ 2},t_{ 2}\big|\big|\Psi\rangle=\Psi(\alpha_{ 1},t_{ 1};a_{ 2},t_{2})$.

Before concluding this section, it is worth pointing out that $||\Psi\rangle$ can be grouped into two orthogonal contributions, the one that describes the happening of the event and the other that refers to the absence of the event:
\begin{eqnarray}\label{sep}
\big|\big|\Psi\big\rangle=\big|\big|\Psi_{{\rm event}}\big\rangle+\big|\big|\Psi_{\rm no~event}\big\rangle,
\end{eqnarray}
where, as $\Psi \big(t_{ 1},\alpha_{ 1};t_{ 2},\alpha_{ 2}\big)=0$,
\begin{eqnarray}\label{sepe}
\big|\big|\Psi_{\rm event}\rangle=\sum_{\substack{a_1 \\ t_1\geq t_0}}~\Psi \big(t_{ 1},\alpha_{ 1}\big)~\big|\big|t_{1},\alpha_{ 1}\big\rangle
\end{eqnarray}
being the state~(\ref{state1}) proposed in the introduction, and
\begin{eqnarray}\label{sepen}
\big|\big|\Psi_{\rm no~ event}\rangle=\sum_{ t_1\geq t_0}~\Psi \big(t_{ 1},0\big)~\big|\big|t_{ 1},0\big\rangle~+~\sum_{ t_2> t_1}~\sum_{\substack{\alpha_1 \\ t_1> t_0}}~\Psi \big(t_{1},\alpha_{ 1};t_{2},0\big) ~\big|\big|t_{ 1},\alpha_{ 1}; t_{ 2},0\big\rangle.
\end{eqnarray}
Notice that by performing position measurements, i.e., by setting $ | \alpha_{ 1} \rangle = | x \rangle $ in Eq.~(\ref{sepe}), we have a space-time-symmetric description of a single event. By keeping in mind the procedure to obtain Eq.~(\ref{PsiGGf}), we can readily extend it for the case of multiple events.

\subsection{Causally connected events}
\label{timelike}

In the last section, we used the conditioned state for a single event [see Eq.~(\ref{Psi})] to obtain Eq.~(\ref{PsiGGf}) via Eq.~(\ref{PWsol}). Here, we will use this procedure and the conditioned state for $2$ events [see Eq.~(\ref{infNN})] to help us to understand the generalization of $||\Psi\rangle$ to the case of $\cal N$ events causally connected. By considering $\cal N$ consecutive measurements on the same system $ \cal S $, it is readily verified that the state~(\ref{PsiGGf}) is extended to  
\begin{eqnarray}\label{PsiGf}
\big|\big|\Psi\big\rangle&=&\sum_{\substack{a_1 \\ t_1\geq t_0}}\Psi \big(t_{ 1},a_{ 1}\big)~\big|\big|t_{ 1},a_{ 1}\big\rangle\nonumber\\
&~&\nonumber\\
&+&\sum_{\substack{a_2 \\ t_2> t_1}}~\sum_{\substack{\alpha_1 \\ t_1> t_0}}\Psi \big(t_{ 1},\alpha_{ 1};t_{ 2},a_{ 2}\big) ~\big|\big|t_{ 1},\alpha_{ 1};t_{ 2},a_{ 2}\big\rangle\nonumber\\
&\vdots&\nonumber\\
&+&\sum_{\substack{a_{\cal N} \\ t_{\cal N}> t_{{\cal N}-1}}}\hdots\sum_{\substack{\alpha_2 \\ t_2> t_1}}~\sum_{\substack{\alpha_1 \\ t_1>t_0}}\Psi \big(t_{1},\alpha_{ 1};t_{ 2},\alpha_{2};\hdots;t_{ {\cal N}},a_{ {\cal N}}\big)~\big|\big|t_{ 1},\alpha_{ 1};t_{ 2},\alpha_{ 2};\hdots;t_{{\cal N}},a_{ {\cal N}}\big\rangle\nonumber\\
&~&\nonumber\\
&+&\sum_{\substack{a_{{\cal N}+1}\\ t_{{\cal N}+1}> t_{{\cal N}}}}~\sum_{\substack{\alpha_n \\ t_{\cal N}> t_{{\cal N}-1}}}\hdots\sum_{\substack{\alpha_2 \\ t_2> t_1}}~\sum_{\substack{\alpha_1 \\ t_1>t_0}}\Psi \big(t_{ 1},\alpha_{ 1};t_{ 2},\alpha_{ 2};\hdots;t_{{\cal N}},\alpha_{ {\cal N}};t_{ {\cal N}+1},a_{ {\cal N}+1}\big)~\big|\big|t_{ 1},\alpha_{ 1};t_{\s 2},\alpha_{2};\hdots;t_{ {\cal N}},\alpha_{ {\cal N}};t_{ {\cal N}+1},a_{ {\cal N}+1}\big\rangle.\nonumber\\
\end{eqnarray}
Here, the first line refers to the first event, the second line to the second event, and so on. Note that the wave functions related to the first event are the same as that of Eq.~(\ref{PsiGGf}), with $\Psi \big(t_{ 1},\alpha_1\big)=\psi_{ S} \big(\alpha_{ 1}|t_{ 1}\big)\chi_1(t_{ 1})$ [with $\chi_{ 1}(t_{ 1})=\chi(t_{ 1})$] and $ \Psi(t_1, 0) =\Gamma_1(t_1, t_0)$ [$\Gamma_1(t_1, t_0)=\Gamma(t_1, t_0)$]. On the other hand, differently from Eq.~(\ref{PsiGGf}), the wave functions associated with the second event can be obtained from the third and fourth terms of Eq.~(\ref{infNN}), which are
\begin{eqnarray}\label{Psic1}
\Psi \big(t_{ 1},\alpha_{ 1};t_{ 2},0\big)=\Gamma_2(t_{2},t_{ 1})~\Psi \big(t_{ 1},\alpha_{ 1}\big)~~~~~~{\rm and}~~~~~\Psi \big(t_{ 1},\alpha_{ 1};t_{ 2},\alpha_{ 2}\big)=\psi_{ S} \big(\alpha_{ 2}|t_{ 2};t_{ 1},\alpha_{ 1}\big)\chi_{ 2}(t_{ 2}|t_{ 1})~\Psi \big(t_{1},\alpha_1\big),
\end{eqnarray}
with $\psi_{S} (\alpha_{ 2}|t_{2};t_{ 1},\alpha_{ 1})$ and $\chi_{ 2}(t_{ 2}|t_{ 1})$ given by Eq.~(\ref{dP2}). The wave function of the second event $\Psi \big(t_{ 1},\alpha_{ 1};t_{ 2},\alpha_{ 2}\big)$ is the joint probability amplitude of $\cal S$ having been in the state $|\alpha_{ 1}\rangle$ with the clock reading $t_{ 1}$, and being in $|\alpha_{ 2}\rangle$ and the clock reading $|t_{2}\rangle$. Besides, the wave function $\Psi \big(t_{ 1},\alpha_{ 1};t_{ 2},0\big)$ is the probability amplitude of $ \cal S $ having been $ | \alpha_1 \rangle_{\cal S} $ with $ \cal C $ reading $ t_{ 1}$, multiplied by the amplitude of the second event not happening in the clock reading interval $] t_{ 1}, t_{ 2}] $. 

At this point, it is important to remember that in QM, it is common to avoid using expressions such as ``probability of the system \emph{being}'' because the system can be in a superposition of states at the instant immediately before measurement. Related to this concern, there is the measurement problem, which refers to the presence of superposition of states at a given instant of time, whereas in our daily life, we observe definite states. In contrast, notice that here we do not speak about the state of the system at a specific moment, but rather as a temporal superposition of an observation. In this broader picture, where the system alone has no meaning, the wave functions actually refer to the probability amplitude of this system \emph{being} in some state at a particular time from the perspective of an observer. Under these circumstances, the measurement problem no longer makes sense.

Let us analyse the event $e$, where $\Psi (t_{ 1},\alpha_{ 1};t_{ 2},\alpha_{ 2};\hdots;t_{ e},a_{e})=\langle t_{ 1},\alpha_{ 1};t_{ 2},\alpha_{ 2};\hdots;t_{ e},a_{ e}||\Psi\rangle$. Note that on each line in Eq.~(\ref{PsiGf}), $ a_{ e} $ appears only in the last label of both the wave function and its respective quantum state, indicating the non-occurrence ($ a_{ e} =0$) and occurrence ($ a_{ e} =\alpha_e)$ of the $e$-th event. Thus, for $ a_{ e} =\alpha_{e}$, the quantum state of the $e$-th event is
\begin{eqnarray}\label{PsiGG1}
&\big|\big|t_{ 1}&,\alpha_{ 1};\hdots;t_{ e-1},\alpha_{ e- 1};t_{e},\alpha_{ e}\big\rangle = \big|t_{e},\alpha_{ e}\big\rangle_{\cal CS}~\big|t_{ 1},\alpha_{\s 1}\big\rangle_{{\cal T}_1{\cal D}_1}~\hdots~\big|t_{ e- 1},\alpha_{ e- 1}\big\rangle_{{\cal T}_{e- 1}{\cal D}_{e-1}}~\big|t_{ e},\alpha_{ e}\big\rangle_{{\cal T}_{e}{\cal D}_{e}}~\hdots~\big|t_{e},0\big\rangle_{{\cal T}_{{\cal N}+1}{\cal D}_{{\cal N}+1}},\nonumber\\
\end{eqnarray}
where the agreement between the states of the clock$+$system and timer$+$detector happens with $ \cal CS $ and $ {\cal T}_{ e} {\cal D}_{ e} $. Thus, as expected, the event associated with the line $e$ is the pair $(t_{ e},\alpha_{ e})$, and, since $t_{ e}>t_{ e- 1}>\hdots>t_{ 1}$, the events $(t_{ 1},\alpha_{ 1};t_{ 2},\alpha_{ 2};\hdots;t_{ e- 1},\alpha_{e- 1})$ are memories recorded in the set $\cal TD$. In addition, for $a_{ e}=0$, we have
\begin{eqnarray}\label{PsiGG2}
\big|\big|t_{ 1},\alpha_{ 1};\hdots;t_{ e- 1},\alpha_{ e- 1};t_{ e},0\big\rangle&=&\big|t_{ e}\big \rangle_{\cal C}~{\hat U}_{\cal S}(t_{ e},t_{ e- 1})\big|\alpha_{e- 1}\big\rangle_{\cal S}\nonumber\\
&\otimes&\big|t_{ 1},\alpha_{ 1}\big\rangle_{{\cal T}_1{\cal D}_1}~\hdots~\big|t_{ e- 1},\alpha_{e- 1}\big\rangle_{{\cal T}_{e-1}{\cal D}_{e-1}}~\big|t_{ e},0\big\rangle_{{\cal T}_{e}{\cal D}_{e}}~\hdots~\big|t_{e},0\big\rangle_{{\cal T}_{{\cal N}+1}{\cal D}_{{\cal N}+1}},
\end{eqnarray}
which depicts the ignorance of $ {\cal D}_{e} $ about $ \cal S $, with $ \cal C $ reading $ t_{e} $. In this manner, this state represents the non-occurrence of the event $e$ in the interval $] t_{e- 1}, t_{ e}] $, and the memory of the events before $ t_{ e} $.

Turning our attention to the wave functions, for $a_{e}=\alpha_{ e}$, the probability amplitude of $(t_{ e},\alpha_{ e})$ with memories $(t_{ 1},\alpha_{ 1})\hdots (t_{ e-1},\alpha_{ e- 1})$ is
\begin{eqnarray}\label{PsiGc}
\Psi \big(t_{ 1},\alpha_{ 1};\hdots;t_{ e},\alpha_{ e}\big)=\psi_{S}(\alpha_{ e}|t_{ e};t_{ e- 1},\alpha_{ e- 1})~\chi_{ e}(t_{ e}|t_{ e-  1})~\Psi \big(t_{ 1},\alpha_{\s 1};\hdots;t_{ e- 1},\alpha_{ e-1}\big),
\end{eqnarray}
where $\psi_{S}(\alpha_{e}|t_{ e};t_{ e- 1},\alpha_{e- 1})={_{\cal S}\big \langle}\alpha_{ e}\big |{\hat U}(t_{e},t_{ e- 1})\big |\alpha_{ e- 1}\big \rangle_{\cal S}$. Lastly, for $a_{e}=0$,
\begin{eqnarray}\label{PsiGc}
\Psi \big(t_{ 1},\alpha_{ 1};\hdots;t_{ e},0\big)=\Gamma_e(t_{e},t_{ e- 1})~\Psi \big(t_{ 1},\alpha_{ 1};\hdots;t_{ e- 1},\alpha_{e- 1}\big)
\end{eqnarray}
is the probability amplitude of the memories $ (t_{ 1}, \alpha_{ 1}) \hdots (t_{e- 1}, \alpha_{e- 1}) $, multiplied by the probability amplitude of the $ e $-th event not having occurred within $]t_{e- 1}, t_{e}] $. Furthermore, by the very formulation of the problem, each event $e$ has normalization condition given by
\begin{eqnarray}\label{normf}
\sum_{\substack{\alpha_e \\ t_e> t_{e-1}}}\hdots~\sum_{\substack{\alpha_1 \\ t_1>t_0}}~\Big|\Psi \big(t_{\s 1},\alpha_{\s 1};\hdots;t_{e},\alpha_{ e}\big) \Big|^2=1.
\end{eqnarray}

To conclude the description of causally connected events, consider that the detector registers more than one degree of freedom of the system, for example, an observable $ \hat \alpha $ of the particle and its position $ \hat {\vec x} $. Thus, we can predict, for instance, the event ``a particle with spin up in the region $ d^3x $ around the position ${\vec x}$ and in the interval $ [t, t + dt] $.'' Rewriting Eq.~(\ref{PsiGf}) more compactly, now we have
\begin{eqnarray}\label{PsiT}
\big|\big|\Psi\big\rangle=\mathlarger{\sum}_{e=1}^{{\cal N}+1}~~\mathlarger{\sum}_{x^\mu_1,\alpha_1}^{x^0_1<x^0_2}~\hdots \mathlarger{\sum}_{x^\mu_{e-1},\alpha_{e-1}}^{x^0_{e-1}< x^0_{e}}~~\mathlarger{\sum}_{x^\mu_{e},a_{e}}~~\Psi \big(x^\mu_{ 1},\alpha_{1};\hdots;x^\mu_{ {e- 1}},\alpha_{ e- 1};x^\mu_{ e},a_{ e}\big)~~\big|\big|x^\mu_{ 1},\alpha_{ 1};\hdots;x^\mu_{ e- 1},\alpha_{e- 1};x^\mu_{e},a_{e}\big\rangle,
\end{eqnarray}
where $\mu=0,1,2,3$, with $x^0=t$, $x^1=x$, $x^2=y$, and $x^3=z$. Notice that time is treated on an equal footing with position and any other physical observable. Equation~(\ref{PsiT}) will be particularly important when we generalize the formalism to an arbitrary number of causally and non-causally connected events.

Finally, it is worth analysing the notation defined above for the case where the environment develops the role of the set of timers. Let us focus only on the second event since the generalization for the $e$-th event is straightforward. For instance, the state of the environment for the situation where the first event happened with the clock reading $ t_{1\s  (k)} $ and the second event has not happened in the interval $] t_{1\s  (k)}, t_{2\s  (\ell)}] $ is
\begin{eqnarray}\label{mevente}
\big|\big|t_{1\s  (k)},\alpha_{ 1}~;~t_{2\s (\ell)},0\big \rangle \equiv \big|t_{2\s (\ell)}\big\rangle_{\cal C} {\hat U}_{\cal S}(t_{2\s  (\ell)},t_{1\s  (k)})~\big|&\alpha_1\big\rangle_{\cal S}&~ \big|\alpha_{ 1}\big\rangle_{{\cal D}_1}~\big|0_{\s (1)},\hdots,\alpha_{1\s  (k)},\hdots,\alpha_{1\s  (\ell)},r_{\s (\ell+1)},\hdots\big\rangle_{{\cal E}_1}\nonumber\\
&\otimes&~\big|0\big\rangle_{{\cal D}_2}~\big|0_{\s (1)},\hdots,0_{\s (k)},\hdots,0_{\s (\ell)},r_{\s (\ell+1)},\hdots\big\rangle_{{\cal E}_2}.
\end{eqnarray}
Note that $ {{\cal E}_{2\s  (\ell)}} $ is the subsystem of the environment that is in ``contact'' with $ {{\cal D}_{ 2}} $ at $ t_{2\s  (\ell)} $ registering this detector in its initial state $ | 0 \rangle_{{\cal D}_2} $. On the other hand, the state of the second event happening at $ t_{2 \s  (\ell)} $, with a memory from $ t_{1 \s  (k)} $, is
\begin{eqnarray}\label{mevente1}
\big|\big|t_{1\s  (k)},\alpha_{ 1}~;~t_{2 \s  (\ell)},\alpha_{ 2}\big \rangle =\big|t_{2 \s (\ell)},\alpha_{ 2}&\big\rangle_{\cal CS} &~\big|\alpha_{ 1}\big\rangle_{{\cal D}_1}~\big|0_{\s (1)},\hdots,\alpha_{1 \s  (k)},\hdots,\alpha_{1 \s  (\ell)},r_{\s (\ell+1)},\hdots\big\rangle_{{\cal E}_1}\nonumber\\
&\otimes&~\big|\alpha_{ 2}\big\rangle_{{\cal D}_2}~\big|0_{\s (1)},\hdots,0_{\s (k)},\hdots,\alpha_{2 \s  (\ell)},r_{\s (\ell+1)},\hdots\big\rangle_{{\cal E}_2}.
\end{eqnarray}
As discussed in Sec.~\ref{env}, from Eqs.~(\ref{mevente}) and~(\ref{mevente1}), we also verify that the arrangement in which the environment stores information about $ \cal S $ determines the readings of the clock $ \cal C $ at the moment of the events. In this more fundamental scenario, instead of referring to the clock reading, an event can also be characterized by both the information stored in the observer, and the internal arrangement used to record this information. Finally, it worth remarking that a relevant situation occurs when an event arises from the acquisition of information by the observer about its surrounding environment, as happens with our perceptions of the external world. With the formalism of this section in mind, now we will discuss the emergence of time based on causally connected events.

\subsection{Time emerging from $||\Psi\rangle$}
\label{time}

As we discussed in the introduction, the assumption that time simply arises from the correlation between the rest of the universe $\cal R$ and the clock $\cal C$ yields serious disagreements with our personal experiences. Thus, although for the sake of comprehension we referred to $ \beta $ and $ t_e$ as measures of time, we will now disregard such a relation. Instead, we will investigate the possibility of time emerging from $ || \Psi \rangle $ by associating the nature of an instant of time with a single event. In this context, as the existence of an event requires a physical object to store information about other systems, time should also depend on this gain of information to exist. From this interpretation, the flow of time will emerge from an asymmetric sequence of events of $||\Psi\rangle$ associated with the same observer. Let us discuss this standpoint carefully.

First, by assuming that an instant of time arises from a single event, the flow of time is related to events of the same observer that occur with probability $1$. This feature implies that the universe from the point of view of this observer (e.g., the set $\cal TD$ or $\cal DE$) is characterized by the existence of all these events. In this manner, for these observations to produce a flow of preferred direction is needed specific asymmetries along a particular sequence of the events in such a way that such asymmetries give rise to the concept of ``passage'' and ``motion'' for the observer. Hence, besides the probability of each event being $1$, the flow should emerge from a sequence of events such as that of Eq.~(\ref{PsiGf}), where the occurrence of a given $(t,\alpha)$ means that the other events with lower values of $t$ (not time) also happens. Note that, without referring to time, the initial condition and the Hamiltonian can generate these events with a well-defined sequence along the growth of $t$. As $t$ increases, these events follow the causal-like sequence described in the introduction: without any reference to time, in Eq.~(\ref{PsiGf}), the constraint ${\hat H}||\Psi\rangle=0$ ``selects'' events such that if $(t_e,\alpha_e)$ is not selected, $(t_{e+1},\alpha_{e+1})$ cannot be selected. This behaviour implies that, along the growth of $t$, there may be a gradual increase of the observer's memory. Under such circumstances, one can say that the causal-like asymmetry along a particular sequence of events (the one with probability $1$ and growing memory) associated with the same observer gives rise to a ``flow'' of events (or time) from the observer's point of view. It should be noted that in this approach the $\beta$-reversal symmetry (not time) of QM still holds.

From the discussion above, for the events of Eq.~(\ref{PsiGf}) to bring about the flow of time, the observer must be able to register information about its surrounding in different degrees of freedom, like the environment or the set $\cal TD$. Thus, the states of distinct events are orthogonal to each other [see Eq.~(\ref{coef})] and can be normalized separately as in Eq.~(\ref{normf}). As a result, these features provide a one-to-one relationship between the possible readings of $\cal C$ and distinguishable states of the observer at the moment of the event. Thus, each event can occur for $\beta$ ranging from $-\infty$ to $\infty$ [with $\chi=0$ for $\beta=t<t_0$], giving rise to a single instant of time from the observer's perspective. This characteristic is similar to the position $x$ of a particle, which has a definite value within $-\infty<x<\infty$ only when it is measured. Besides, since the arrangement employed to store information has longer memories for higher values of the clock reading, the growth of $\beta=t$ works as a good track of the flow of time.

It is readily seen that the proposal above is coherent with our personal experience of time. Here, one should consider our brain playing a role similar to that of the set of $\cal TD$ or $\cal E$, so that we perceive all events within $||\Psi\rangle$ associated with our brain. Now, by taking $\cal E$ as the observed system, the events become our perceptions that come from the acquisition of information stored by our brain about the universe around us. In this context, similarly to the conclusion above, the sequence of causal-like events contained in $ || \Psi \rangle $, which follows the increase of memory, is responsible for our perception of the flow of time. At this point, we verify the epistemic asymmetry of time, where the past is remembered, whereas the future is inferred. Nevertheless, it should be emphasized that the asymmetric sequence of events are properties of QM itself, and thus the flow of time can be, in principle, associated with any observer with many degrees of freedom, not necessarily a conscious being. It is worth remarking that this approach is compatible with works in philosophy and
neuroscience of consciousness concerning our perception of motion, which takes place in discrete processing frames
or snapshots~\cite{Van}.

It is worth pointing out that from the perspective of the environment $\cal E$ (see Sec.~\ref{env}), the smallest time interval is not given by the distance $(t_{ e+1}-t_{e})$ between two consecutive measurements of $\cal S$. Remember that at every step $\delta t_{\cal DE}$, the environment records information about the detector ($|0\rangle_{\cal D}$ or $|\alpha\rangle_{\cal D}$), and thus a new event (or instant of time) takes place from the environment's point of view. Under these circumstances, the events associated with the highest frequency measurements (observations of $\cal D$) work as a background time for those events with lower frequency measurements (observations of $\cal S$). On this basis, e.g. in Eq.~(\ref{mevente1}), $k\delta t_{\cal DE}=(t_{(\s k)}-t_0)$ is the time interval (or the number of measurements on $\cal D$) for $\cal E$ to observe $\cal S$ with a well-defined $\alpha$.

We conclude this section by highlighting that in this manuscript, we abdicated the concept of an external parameter $t$ that flows generating events and motion (Newtonian time), as well as the interpretation of time arising from correlations (PW time). Instead, here, we analyse the viewpoint in which time and its flow emerge from an asymmetric sequence of orthogonal events within $||\Psi\rangle$ characterized by the same observer.

\subsection{Arbitrary events}
\label{spacelike}

As mentioned earlier, the generalization of Eq.~(\ref{PsiT}) for cases where $\cal S$ is composed of entangled systems is the subject of a work in progress. Thus, in this section, we still deal with uncorrelated measured systems but assuming noncausally connected events. To this end, be $ \cal M $ independent systems with each one of them submitted to ${\cal N}_{s} -1 $ consecutive measurements, with $ s = 1, 2, \hdots, {\cal M} $. The measurements between different values of $s$ are independent of each other, and thus the non-causal (and spacelike) events are identified by the set $\{s\}$, so that each event is characterized by the pair $ (x^\mu_{ se_s}, \alpha_{ se_s})$, where $ e_s = 1, 2, \hdots, {\cal N}_{s}$ represent the causally related events for a given $s$. Under these circumstances, as each measurement can be of a different observable, Eq.~(\ref{PsiT}) becomes
\begin{eqnarray}\label{PsiTS}
\big|\big|\Psi\big\rangle=\mathlarger{\sum}_{s=1}^{ \cal M}~\mathlarger{\sum}_{e_s=1}^{{\cal N}_{s} }~~\mathlarger{\sum}_{x_{ s1}^{\mu},\alpha_{ s1}}^{\s x^0_{ s1}<\s x^0_{ s2}}\hdots~\mathlarger{\sum}_{x_{ se_s}^{\mu},a_{ se_s}}~\Psi \left(\big\{x_{ s  1}^{\mu},\alpha_{ s 1};\hdots;x^{\mu}_{ s e_s},a_{ s e_s}\big\}_{s=1}^{\cal   M}\right)~~ \bigotimes_{s=1}^{\cal  M}~\big|\big|x^{\mu}_{ s 1},a_{ s 1};\hdots;x^{\mu}_{ se_s},a_{se_s}\big\rangle_s~,
\end{eqnarray}
where $\{x_{ s 1}^{\mu},\alpha_{s  1};\hdots;x^{\mu}_{ s e_s},a_{ s e}\big\}_{s=1}^{\cal  M}=\{\{x_{ 1 1}^{\mu},\alpha_{ 1 1};\hdots;x^{\mu}_{{1} e_s},a_{{ 1} e_s}\big\},\hdots,\{x_{ {\cal M} 1}^{\mu},\alpha_{ {\cal M}  1};\hdots;x^{\mu}_{ { {\cal M}} e_s},a_{ { {\cal M}}e_s}\big\}\}$, and the function $\Psi(...)$ is separable for different values of $s$. A more detailed discussion of this state will be carried out in future works. In Eq.~(\ref{PsiTS}), for convenience, we isolate causally connected events in the same ket $ || \rangle $. Notice that, in contrast to the causal events for a given $s$, where always $t_{ se_s}>t_{ se_s- 1}$,  for different values of $ s $, there is no correlation between the times $ t_{se_s} $. Moreover, note that in Eq.~(\ref{PsiTS}), for fixed values of $ e_s $ and $ s $, we have the contribution of $ e_s $ sums ($s1, s2, \hdots, se_s $). The reason for this is that for a given event $e_s$, we have $ e_s-1 $ memories referring to previous events that must occur for the event $e_s$ to happen.

\section{Conclusion}
\label{conclusion}

First, this work proposed a quantum formalism that treats time on an equal footing with any other observable. To this end, as relativity deals with events describing space and time symmetrically, we extended the classical concept of events to the quantum realm by defining an event as a transfer of information between physical systems. Following the Page and Wootters formalism, we described events by assuming that the rest of the universe $ \cal R $ encompasses the system of interest $ \cal S $, a detector $ \cal D $, and, at first, a timer $ \cal T $ synchronized with a non-interacting clock $ \cal C $. By stopping its counting,  $\cal T$ could register the reading of the clock $ \beta = t $ at the moment at which the detector measures an observable $\hat \alpha$ of $ \cal S $. The recording of information about $ \cal CS $ within $ \cal TD $ defined the single event $ (x^\mu, \alpha) $, which was described by a joint probability amplitude $ \Psi ( x^\mu, \alpha) $ and represented by the state $||x^\mu,\alpha\rangle$.

We calculated how the probabilities of a given observable can be affected by the detection time distribution when the measurement process is not ``instantaneous.'' Then, we extended this approach for an arbitrary number of causally connected events, in which we obtained a global state represented by a superposition of the detection times of all possible events predicted by a Wheeler-DeWitt equation. The treatment for non-causal events was also briefly addressed. In addition, we verified that a macroscopic observer could play the same role as a set of timers by registering information in its degrees of freedom through many different orthogonal arrangements.

Our second goal was to obtain the nature of time from the formalism of events. To that end, we first disassociated the clock observable with the concept of time, and then assumed that an instant of time emerges from a single event. This assumption was made possible by the observer's ability to store information within distinguishable states, which also provided a one-to-one relationship between the reading of a non-interacting clock and arrangement of the observer's state at the event. In this way, the flow of time could emerge from the observer's perspective via a causal-like sequence of events associated with this observer.

A reason for this interpretation of the flow of time came from the fact that $||\Psi\rangle$ predicts each event with probability 1, and so all these events were assumed to be ``real'' from the observer's perspective. Besides that, the occurrence of a given event depended on the happening of other ones with lower values of the clock reading, which allowed the increase of the observer's memory along the growth of $t$. From our conscious perception standpoint, the passage of time was approached as our experience of all causal-like events (and their memories) within $||\Psi\rangle$ related to our brain.

\begin{acknowledgments}
I thank Prof. Vlatko Vedral and his group, Frontiers of Quantum Physics, for the fruitful discussions that contributed significantly to the development of this work. I acknowledge financial support from Coordena\c{c}\~ao de Aperfei\c{c}oamento de Pessoal de
N\'ivel Superior (CAPES) through its program GPCT - 17/2016 (Grant 88887.312745/2018-00), and Conselho Nacional de Desenvolvimento Cient\'ifico e Tecnol\'ogico (CNPq) through its program 09/2020 (Grant 311043/2017-8).
\end{acknowledgments}

\appendix*
\label{ap}
\section{Calculation of $\big|\psi(t_{(1)}\big\rangle_{\cal R}$}

In this work, as we do not intend to solve Eq.~(\ref{Schr}) for a specific Hamiltonian ${\hat H}_{\cal R}={\hat H}_{\cal STD}$~(\ref{hamilt}), the evolved state $|\psi(t_{(N)})\rangle_{\cal R}$~(\ref{infN}) should represent the measurement process introduced in Sec.~\ref{form1e} as generally as possible. Therefore, let us verify that Eq.~(\ref{inf1}) can describe, for instance, a measurement in which the detector interacts differently with distinct states $|\alpha\rangle_{\cal S}$, i.e.,
\begin{eqnarray}
\label{inf1a}
\big|\psi(t_{\s (1)})\big\rangle_{\cal R}&=&\bigg[\sum_\alpha~\sqrt{1-\delta p_{\alpha \s(1)}}~{\rm e}^{i\varphi_{\alpha(1)}}~{\hat M}_\alpha~{\hat U}_{\cal S}(t_{\s (1)},t_0)\bigg]\big|0\big\rangle_{\cal S}~\big|t_{\s (1)}\big\rangle_{\cal T}~\big|0\big\rangle_{\cal D}\nonumber\\
&+&\bigg[\sum_\alpha\sqrt{\delta p_{\alpha\s (1)}}~{\rm e}^{i{\tilde \varphi}_{\alpha(1)}}~{\hat M}_\alpha~{\hat U}_{\cal S}(t_{\s (1)},t_0)\bigg]\big|0\big\rangle_{\cal S}~\big|t_{\s (1)}\big\rangle_{\cal T}~\big|\alpha\big\rangle_{\cal D},
\end{eqnarray}
where $\delta p_{\alpha \s(1)}\ll 1$ is a probability related to the measurement of $|\alpha\rangle_{\cal S}$, $\varphi_{\alpha(1)}$ and ${\tilde \varphi}_{\alpha(1)}$ are phases associated with the unmeasured and measured situations respectively, and ${\hat U}_{\cal S}(t_{\s (1)},t_0)=\exp\{-i{\hat H}_{\cal S}(t_{\s (1)}-t_0)/\hbar\}$. It is worth mentioning that rewriting Eq.~(\ref{inf1a}) in the form of Eq.~(\ref{inf1}) is extremely useful for clarity and understanding of the formalism of events that we present in this manuscript. To this end, we have $\delta p_{\s(1)}$ of Eq.~(\ref{inf1}) defined as
\begin{eqnarray}
\label{inf1b}
\delta p_{\s(1)}=\sum_\alpha~\delta p_{\alpha\s (1)}~{\big |} _{\cal S}{\big\langle}\alpha \big|{\hat U}_{\cal S}(t_{\s (1)},t_0)\big|0\big {\rangle_{\cal S}}\big|^2,
\end{eqnarray}
which is the probability of the measurement taking place in the interval $[t_0,t_{\s(1)}]$ regardless of the outcome $\alpha$. Thus, Eq.~(\ref{inf1a}) can be written in the form of Eq.~(\ref{inf1}) by defining
\begin{eqnarray}
\label{inf1c}
{\hat U}_{\cal S}^{ \s 0(1)}(t_{\s (1)},t_0)=\sum_\alpha~\sqrt{\frac{1-\delta p_{\alpha \s(1)}}{{1-\delta p_{\s(1)}}}}~{\rm e}^{i(\varphi_{\alpha(1)}-\varphi_{(1)})}~{\hat M}_\alpha~{\hat U}_{\cal S}(t_{\s (1)},t_0)
\end{eqnarray}
and
\begin{eqnarray}
\label{inf1d}
_{\cal S}{\big\langle}\alpha|{\hat U}_{\cal S}^{ \s m(1)}(t_{\s (1)},t_0)|0\big {\rangle_{\cal S}}=\sqrt{\frac{\delta p_{\alpha \s(1)}}{{\delta p_{\s(1)}}}}~{\rm e}^{i{\tilde \varphi}_{\alpha(1)}}~_{\cal S}{\big\langle}\alpha|{\hat U}_{\cal S}(t_{\s (1)},t_0)|0\big {\rangle_{\cal S}}.
\end{eqnarray}
From Eqs.~(\ref{inf1c}) and~(\ref{inf1d}), we observe that for each time step of the evolution, the operators ${\hat U}_{\cal S}^{ \s 0(1)}(t_{\s (1)},t_0)$ and ${\hat U}_{\cal S}^{ \s m(1)}(t_{\s (1)},t_0)$ must be updated. Nevertheless, notice that if the detector interacts equally with all states $|\alpha\rangle_{\cal S}$, $\delta p_{\alpha\s(1)}=\delta p_{\s(1)}$, and thus we can have ${\hat U}_{\cal S}^{ \s 0(1)}(t_{\s (1)},t_0)={\hat U}_{\cal S}^{ \s m(1)}(t_{\s (1)},t_0)={\hat U}_{\cal S}(t_{\s (1)},t_0)$.

\end{document}